\documentclass[preprintnumbers,endnote,nofootinbib,11pt,superscriptaddress]{article}
\pdfoutput=1
\usepackage{epsfig}
\usepackage{epstopdf}
\usepackage[table]{xcolor}

\usepackage[utf8]{inputenc}
\usepackage{graphicx}
\usepackage{subcaption}
\usepackage{amssymb}
\usepackage{amsxtra}
\usepackage{amsmath}
\usepackage{booktabs,multirow,tabularx}
\usepackage{slashed}
\usepackage{float}
\usepackage{placeins}
\usepackage{rotating}
\usepackage{lscape}
\usepackage{color}
\usepackage{hyperref}
\usepackage[compress,numbers,sort]{natbib}
\usepackage{enumitem}
\usepackage[normalem]{ulem}

\usepackage[Q=yes,pverb-linebreak=no]{examplep}

\usepackage[left=0.14\paperwidth ,right=0.14\paperwidth]{geometry}

\makeatletter
\newcommand{\specialcell}[1]{\ifmeasuring@#1\else\omit$\displaystyle#1$\ignorespaces\fi}
\makeatother

\DeclareGraphicsExtensions{.eps}
\graphicspath{{./}}
\newcommand{\pt}{p_{\scriptscriptstyle T}}

\newcommand{\be}{\begin{equation}}
\newcommand{\ee}{\end{equation}}

\newcommand{\mt}{m_t}

\newcommand\sss{\scriptscriptstyle}

\newcommand{\mz}{m_{ \sss Z}}
\newcommand{\mzz}{M_{ \sss ZZ}}

\def\beq{\begin{equation}}
\def\bea{\begin{eqnarray}}
\def\eeq{\end{equation}}
\def\eea{\end{eqnarray}}
\def\beqnl{\begin{align}}
\def\endal{\end{align}}

\newcommand\parts{\hat{s}}
\newcommand\partt{\hat{t}}
\newcommand\partu{\hat{u}}

\newenvironment{appendletterA}
 {
  \typeout{ Starting Appendix \thesection }
 \renewcommand{\thesection}{\Alph{section}}
  \setcounter{section}{0}
  \setcounter{equation}{0}
  
 }{
  \typeout{Appendix done}
 }

 \newenvironment{appendletterB}
 {
  \typeout{ Starting Appendix \thesection }
 \renewcommand{\thesection}{\Alph{section}}
  \setcounter{section}{1}
  \setcounter{equation}{0}
  
 }{
  \typeout{Appendix done}
 }

\makeatletter
\newcommand{\normalorbold}{%
  \ifnum\pdf@strcmp{\math@version}{bold}=\z@ bx\else m\fi
}
\makeatother

\begin{document}\color{black}
\begin{titlepage}
\nopagebreak
{\flushright{
        \begin{minipage}{5cm}
         COMETA-2024-06 \\TTP24-010
        \end{minipage}        }

}

\renewcommand{\thefootnote}{\fnsymbol{footnote}}
\vspace{1cm}
\begin{center}
  {\Large \bf \color{magenta} Virtual QCD corrections to $gg \to ZZ$: top-quark loops from a transverse-momentum expansion}
  
\bigskip\color{black}\vspace{0.6cm}
     {\large\bf
       Giuseppe Degrassi$^a$\footnote{email: giuseppe.degrassi@uniroma3.it},
       Ramona Gr\"{o}ber$^{b}$\footnote{email: ramona.groeber@pd.infn.it},
       Marco Vitti$^{c,d}$\footnote{email: marco.vitti@kit.edu} }
     \\[7mm]     
{\it (a) Dipartimento di Matematica e Fisica, Universit{\`a} di Roma Tre and \\
 INFN, sezione di Roma Tre, I-00146 Rome, Italy}\\[1mm]
{\it (b) Dipartimento di Fisica e Astronomia 'G.~Galilei', Universit\`a di
  Padova and INFN, sezione di Padova, I-35131 Padova, Italy}\\[1mm]
{\it (c) Institute for Theoretical Particle Physics, Karlsruhe Institute of
  Technology (KIT), D-76131 Karlsruhe, Germany}\\[1mm]
{\it (d) Institute for Astroparticle Physics, Karlsruhe Institute of Technology
  (KIT), D-76344 Eggenstein-Leopoldshafen, Germany}

\end{center}

\bigskip
\bigskip
\bigskip
\vspace{0.cm}

\begin{abstract}
 We present the virtual corrections due to the top-quark loops for the
 process $gg \to ZZ$ at next-to-leading order in QCD. The associated
 two-loop box diagrams are computed using a small-transverse-momentum
 expansion. Our results are then merged with those available in the
 complementary energy region, obtained via a high-energy expansion, in order to
 provide an analytic result that is valid in the whole phase space.  The
 results presented allow for an efficient modelling of the signal--background
 interference as well as the irreducible
 background in off-shell Higgs production.

\end{abstract}
\vfill  
\end{titlepage}    

\setcounter{footnote}{0}

\section{Introduction}
\label{sec:introzz}

The quest for the determination of the properties of the Higgs boson,
after its discovery at the LHC, has been very rewarding. While 
the main Higgs production channels have been established and new challenging
decay modes are being studied (see e.g.~refs.~\cite{ATLAS:2022vkf,
  CMS:2022dwd}), a large variety of measurements have entered the precision domain, requiring improved predictions for Higgs-related
processes within the Standard Model (SM)
\cite{Heinrich:2020ybq, Huss:2022ful}. A significant role in this
program is played by the production of a pair of $Z$ bosons, $pp
\rightarrow ZZ$, which has been one of the discovery channels of
the Higgs and today is important both as a probe of the electroweak (EW)
symmetry breaking and for precision Higgs physics \cite{ATLAS:2023zrv, ATLAS:2023dew}.

In the theoretical SM prediction of $pp \rightarrow ZZ$ production two
partonic sub-processes have to be considered. The first one is
quark-antiquark annihilation, $q\bar{q} \rightarrow ZZ$, which gives
the largest contribution to the hadronic cross section. The
leading-order (LO) amplitude for this channel is related to purely EW
tree-level diagrams \cite{PhysRevD.19.922}, and corrections through
next-to-next-to-leading order (NNLO) in QCD
\cite{PhysRevD.43.3626,MELE1991409,Cascioli:2014yka,Gehrmann:2014bfa,Caola:2014iua,Gehrmann:2015ora}
and through next-to-leading order (NLO) in the EW theory
\cite{Accomando:2004de,Bierweiler:2013dja,Baglio:2013toa} are
available.  The second partonic contribution, which is the main topic
of this paper, comes from the gluon-initiated channel, $gg \rightarrow
ZZ$. The gluon-initiated channel at LO is associated to one-loop
diagrams, which contribute as an important $\mathcal{O}(\alpha_s^2) $
correction, accounting for about 10\% of the hadronic cross section at
NNLO for $\sqrt{s}=13$ TeV \cite{Grazzini:2018owa}.

The one-loop diagrams for $gg \rightarrow ZZ$ at LO have been computed
for the first time in refs.~\cite{PhysRevD361570,GLOVER1989561}. They
feature two topologies: the triangles (see fig.~\ref{fig:ggzzLO}(a))
are associated to Higgs production via the sub-process $gg \rightarrow
H \rightarrow ZZ$, while the box diagrams (fig.~\ref{fig:ggzzLO}(b))
are related to the process of non-resonant (a.k.a.~\emph{continuum})
$ZZ$ production, which constitutes an irreducible background in
experimental Higgs searches. Moreover, continuum production plays
a relevant role in the indirect determination of the Higgs total decay
width, $\Gamma_H$ \cite{ATLAS:2018jym,CMS:2019ekd}. Indeed, in
refs.~\cite{Caola:2013yja,Campbell:2013una} it has been suggested that
upper limits on $\Gamma_H$ can be obtained from the investigation of
the invariant-mass distribution of the two $Z$ bosons system, $\mzz$, away from the region where the Higgs
is produced on shell\footnote{Off-shell Higgs production can also
  provide an excellent probe of new physics, for instance of
  light-quark Yukawa couplings \cite{Balzani:2023jas}, modified
  trilinear Higgs self-coupling \cite{Haisch:2021hvy} or Higgs portal
  models \cite{Haisch:2022rkm, Haisch:2023aiz}}
\cite{Kauer:2012hd}. Under the assumption that no beyond-SM physics
spoils the known correlation between on- and off-shell amplitudes
\cite{Englert:2014aca,Englert:2014ffa}, the latest experimental
measurements could exclude, for the
first time, deviations in $\Gamma_H$ below $\mathcal{O}(100 \%)$ from its SM
prediction \cite{CMS:2022ley, ATLAS:2023dnm}.

 In this situation, a good theoretical control over the destructive
 interference between the Higgs-mediated and continuum amplitudes is
 crucial. This motivates the inclusion of NLO QCD corrections to the
 gluon-initiated channel.  The NLO corrections to the Higgs-mediated
 diagrams are known exactly, adapting results for the production of a
 single Higgs with virtuality $\mzz$
 \cite{Anastasiou:2006hc,Aglietti:2006tp,Harlander:2005rq}. Concerning
 the continuum term, exact analytic results for the related two-loop
 box integrals have been obtained in the case of loops of massless
 quarks \cite{vonManteuffel:2015msa,Caola:2015psa}. We remark that, at
 the level of the inclusive $gg \rightarrow ZZ$ cross section, the
 contribution from light quarks is the dominant one, and furthermore
 it constitutes more than 50\% of the $\mathcal{O}(\alpha_s^2)$
 corrections to $ZZ$ production
 \cite{Cascioli:2014yka,Grazzini:2018owa}.

The top-quark contribution to continuum $gg \rightarrow ZZ$ starts to
become relevant for invariant masses  in the
range $\mzz > 2 \mz$
and it is expected to be substantial in the region where $\mzz$ is
larger than twice the mass of the top, $\mt$.
At present, the top-quark contribution is not known in full
analytic form, as the scale associated to the mass of the heavy quark
complicates the calculation of the two-loop box integrals. Several
approximate analytic evaluations of this contribution have been discussed in the literature. The
large-mass expansion (LME) has been used in
refs.~\cite{Melnikov:2015laa,Caola:2016trd,Campbell:2016ivq} to obtain
reliable predictions in the region $2\, \mz < \mzz \lesssim 2\, \mt$.
In ref.~\cite{Campbell:2016ivq} an improvement of the LME
results by means of conformal mapping and Padé approximants has also
been studied. A further improvement of the LME via an expansion around
the top threshold has been presented in
ref.~\cite{Grober:2019kuf}. Analytic predictions that are reliable for
large invariant masses have been obtained using the so-called High-Energy (HE)
expansion \cite{Davies:2020lpf}. Still, the latter approach is
expected to break down in the region $\mzz \lesssim 750$ GeV, and in
ref.~\cite{Davies:2020lpf} Padé approximants were used in order to
improve the expansion. Finally, exact results based on numerical
approaches have been obtained in
refs.~\cite{Agarwal:2020dye,Bronnum-Hansen:2021olh}, showing good
agreement with ref.~\cite{Davies:2020lpf} in the ranges of small or large invariant masses. Very recently, the numerical results of ref.~\cite{Agarwal:2020dye} have been used to compute the full NLO QCD corrections to $gg \to ZZ$ \cite{Agarwal:2024pod}.

In this paper, we present the computation of the contribution from
top-quark loops to the virtual NLO QCD corrections to $gg \rightarrow ZZ$. In the calculation we 
employ an expansion in the transverse momentum of the $Z$, $\pt$,
following refs.~\cite{Bonciani:2018omm,Alasfar:2021ppe}.
Our first goal is to provide an accurate
approximation of the virtual QCD corrections that can be valid in the
invariant-mass region that so far has not been covered by any of the
analytic approaches discussed above, namely the region $350 \lesssim
\mzz \lesssim 750$ GeV.  Since  the $\pt$ expansion 
``contains'' the LME result, the region of validity of our approach
is actually given by  $2\, \mz < \mzz \lesssim 750$ GeV.

Furthermore, it has been previously
shown that an expansion in the forward kinematics  \cite{Bellafronte:2022jmo, Davies:2023vmj}, like the $\pt$ expansion,  and
the HE expansion can be combined in order to approximate the two-loop
box amplitudes with good accuracy over the whole phase space. In the
second part of the paper, then, we show that this can be done also for
$gg \to ZZ$, and we discuss the combination of our new results with
those of ref.~\cite{Davies:2020lpf}.

The paper is organized as follows.  In section~\ref{sec:def} we set
the notation and we comment on the application of the $\pt$ expansion
to the $gg\to ZZ$ case. The next section is devoted to a comparison
between the known LO amplitude and our approximation. In
section~\ref{sec:nlo} we then consider the application of the $\pt$
expansion at NLO and and discuss how our $\pt$-expanded results can be
merged with those derived via the HE expansion. Section~\ref{sec:results} contains our NLO results, and we present our conclusions in section~\ref{sec:concl}. The
paper is complemented by two appendices. In appendix \ref{app:projezz}
we report the explicit expressions for the projectors we
employ in the calculation. We present also the relation between our
form factors and the ones used in ref.~\cite{Davies:2020lpf}. In appendix
\ref{app:analytzz} we report
the exact results for the NLO triangle and the reducible double-triangle
contributions.

\section{Definitions and the $\pt$-expansion method}
\label{sec:def}
\subsection{Definitions}
We consider the process
$g_a^\mu (p_1) g_b^\nu (p_2) \rightarrow Z^\rho (p_3) Z^\sigma (p_4)$.
The amplitude can be defined as 

\begin{equation}
  \mathcal{A} = \sqrt{2} \mz^2 G_F ~ \frac{\alpha_s (\mu_R)}{\pi}
  \delta_{a b}~ \epsilon^a_\mu (p_1) \epsilon^b_\nu (p_2) \epsilon_\rho^* (p_3)
  \epsilon_\sigma^* (p_4) ~ \hat{\mathcal{A}}^{\mu \nu \rho \sigma} (p_1, p_2, p_3),
\end{equation}
where $G_F$ is the Fermi constant, $\alpha_s(\mu_R)$ is the strong
coupling constant evaluated at a renormalisation scale $\mu_R$ and the
polarization vectors of the gluons and the $Z$ bosons are
$\epsilon^a_\mu (p_1), \epsilon^b_\nu (p_2)$ and $\epsilon_\rho (p_3),
\epsilon_\sigma (p_4)$, respectively. The Lorentz structure of the
amplitude is encoded in the tensor
$\hat{\mathcal{A}}^{ \mu \nu \rho \sigma} (p_1, p_2, p_3)$, whose most
general decomposition consists of 138 Lorentz structures. However, by imposing
the transversality of the external polarization vectors w.r.t.~the
relative four-momentum
\begin{equation}
\epsilon(p_i) \cdot p_i = 0 \qquad  i=1,\dots,4,
\label{eq:trasv}
\end{equation}
and by fixing the gauge of the external gluons with 
\begin{equation}
  \epsilon(p_1) \cdot p_2 = 0 \qquad \epsilon(p_2) \cdot p_1 = 0,
  \label{eq:gauge}
\end{equation}
$\hat{\mathcal{A}}^{ \mu \nu \rho \sigma}$ can be written as a
linear combination of 20 Lorentz structures ~\cite{vonManteuffel:2015msa,Caola:2015ila,Davies:2020lpf}
\begin{equation}
  \hat{\mathcal{A}}^{\mu \nu \rho \sigma} (p_1, p_2, p_3) = \sum_{i=1}^{20}
  S_i^{\mu \nu \rho \sigma} f_i (\hat{s},\hat{t},\hat{u}, \mt,\mz),
\label{eq:sideczz}
\end{equation}
where the scalar form factors $f_i$ depend, besides $\mt$ and
$\mz$,  on the partonic Mandelstam
variables. Assuming all momenta to be incoming, the latter are defined
as
\begin{equation}
\hat{s} = (p_1+p_2)^2, \quad \hat{t} = (p_1+p_3)^2, \quad \hat{u} = (p_2+p_3)^2
\end{equation}
and the relation $\hat{s}+\hat{t}+\hat{u}= 2\, \mz^2$ is satisfied. We checked that the Lorentz structures that give a nonzero
contribution to the amplitude are
 \begin{align}
S_{1}^{\mu \nu \rho \sigma} &= g^{\mu \nu } g^{\rho \sigma } & S_{2}^{\mu \nu \rho \sigma}&= g^{\mu \rho } g^{\nu \sigma } & S_{3}^{\mu \nu \rho \sigma}&= g^{\mu \sigma } g^{\nu \rho } & S_{4}^{\mu \nu \rho \sigma}&= p_1^{\rho } p_3^{\nu } g^{\mu \sigma } \nonumber \\
 S_{5}^{\mu \nu \rho \sigma} &= p_2^{\rho }
   p_3^{\nu } g^{\mu \sigma } & S_{6}^{\mu \nu \rho \sigma}&= p_1^{\rho } p_3^{\mu } g^{\nu \sigma } & S_{7}^{\mu \nu \rho \sigma}&= p_2^{\rho } p_3^{\mu } g^{\nu \sigma } & S_{8}^{\mu \nu \rho \sigma}&= p_3^{\mu } p_3^{\nu } g^{\rho \sigma
   } \nonumber \\
 S_{9}^{\mu \nu \rho \sigma}&= p_1^{\rho } p_1^{\sigma } g^{\mu \nu } & S_{10}^{\mu \nu \rho \sigma}&= p_1^{\rho } p_2^{\sigma } g^{\mu \nu } & S_{11}^{\mu \nu \rho \sigma}&= p_1^{\sigma } p_2^{\rho } g^{\mu \nu } & S_{12}^{\mu \nu \rho \sigma}&= p_2^{\rho }
   p_2^{\sigma } g^{\mu \nu } \label{eq:lorstructzz} \\
   S_{13}^{\mu \nu \rho \sigma}&= p_1^{\sigma } p_3^{\nu } g^{\mu \rho } & S_{14}^{\mu \nu \rho \sigma}&= p_2^{\sigma } p_3^{\nu } g^{\mu \rho } & S_{15}^{\mu \nu \rho \sigma}&= p_1^{\sigma } p_3^{\mu } g^{\nu \rho
   } & S_{16}^{\mu \nu \rho \sigma}&= p_2^{\sigma } p_3^{\mu } g^{\nu \rho } \nonumber \\ 
   S_{17}^{\mu \nu \rho \sigma}&= p_1^{\rho } p_1^{\sigma } p_3^{\mu } p_3^{\nu } & S_{18}^{\mu \nu \rho \sigma}&= p_1^{\rho } p_2^{\sigma } p_3^{\mu }
   p_3^{\nu } & S_{19}^{\mu \nu \rho \sigma}&= p_1^{\sigma } p_2^{\rho } p_3^{\mu } p_3^{\nu } & S_{20}^{\mu \nu \rho \sigma}&= p_2^{\rho } p_2^{\sigma } p_3^{\mu } p_3^{\nu } \nonumber
\end{align}
where we follow the numbering of ref.~\cite{Davies:2020lpf}.

In order to simplify the evaluation of the cross section, in our work
we express the amplitude in terms of a set of orthonormal projectors
\begin{equation}
  \hat{\mathcal{A}}^{\mu \nu \rho \sigma} (p_1, p_2, p_3) = \sum_{i=1}^{20}
\mathcal{P}_i^{\mu \nu \rho \sigma} \mathcal{A}_i (\hat{s},\hat{t},\hat{u},\mt,\mz) ,
\label{eq:clproje}
\end{equation}
where the tensors $\mathcal{P}_i^{\mu \nu \rho \sigma}$ are
constructed as linear combinations of the Lorentz structures defined in
eqs.~(\ref{eq:lorstructzz}), using a Gram-Schmidt procedure for the
orthogonalization. In appendix~\ref{app:projezz} we give the explicit
expressions for the projectors. Here we point out that, to efficiently perform
the $\pt$ expansion \cite{Alasfar:2021ppe},
we choose  the projectors to be either symmetric or anti-symmetric under the
interchange $\{\mu \leftrightarrow \nu, p_1 \leftrightarrow p_2\}$.
This choice  also allows to reduce
the number of relevant form factors\footnote{In fact, we observed that
enforcing the additional Bose symmetry $\{\rho \leftrightarrow \sigma,
p_3 \leftrightarrow p_4\}$ further reduces the relevant form factors
to 12. While this may be used for improving the practical
implementation of our results, in this paper we use the 16 form
factors.} from 20 to 16. We present our results in terms of the
$\mathcal{A}_i$ form factors of eq.~(\ref{eq:clproje}), while in
appendix~\ref{app:projezz} we include the relations to obtain the latter
as a combination of the $f_i$ in eqs.~(\ref{eq:sideczz}).

\begin{figure}
\centering
\begin{subfigure}{0.3\textwidth}
  \centering
\hspace*{-0.8cm}  \includegraphics[width=1.\linewidth]{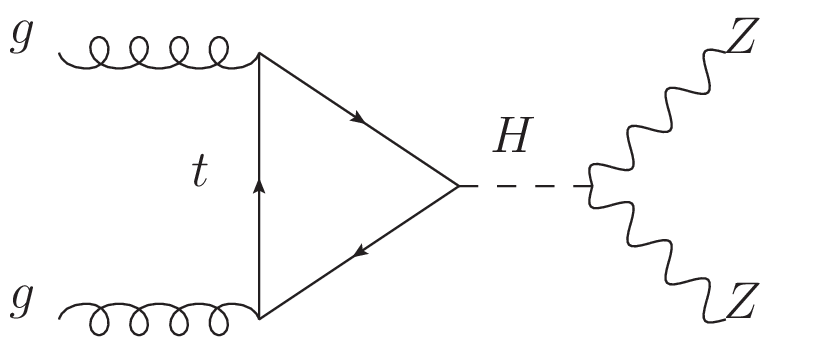}
  \caption{}
\end{subfigure}%
\begin{subfigure}{0.3\textwidth}
  \centering
  \includegraphics[width=1.\linewidth]{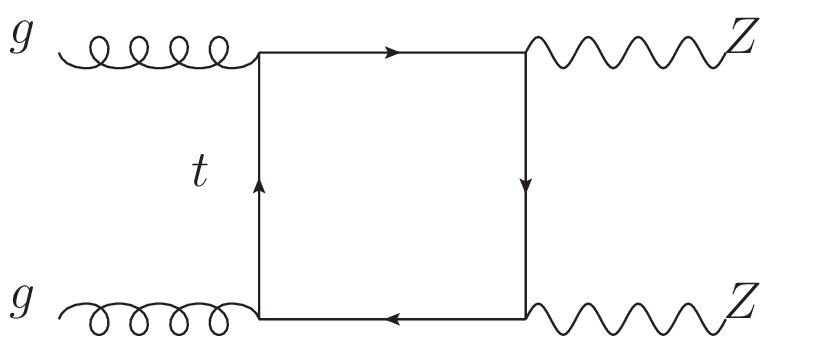}
  \caption{}
\end{subfigure} 
\caption{Representative Feynman diagrams contributing to the
  $gg \rightarrow ZZ$ amplitude at LO. Only the contribution from top-quark
  loops is shown.}
\label{fig:ggzzLO}
\end{figure}

\begin{figure}[t]
\centering
\begin{subfigure}{.3\textwidth}
  \centering
     \includegraphics[width=1.\linewidth]{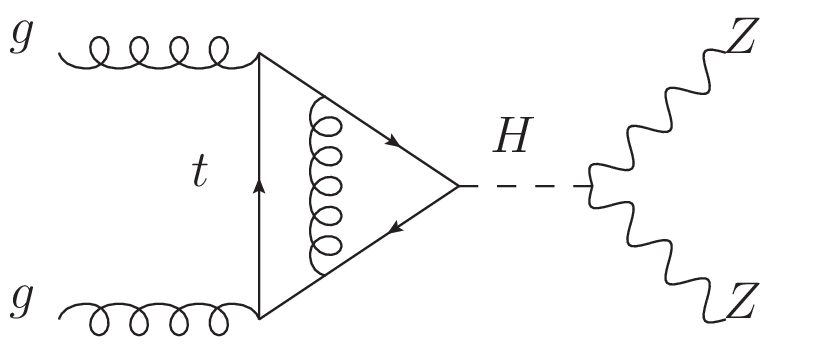}
  \caption{}
\end{subfigure}%
\begin{subfigure}{.3\textwidth}
  \centering
   \includegraphics[width=1.\linewidth]{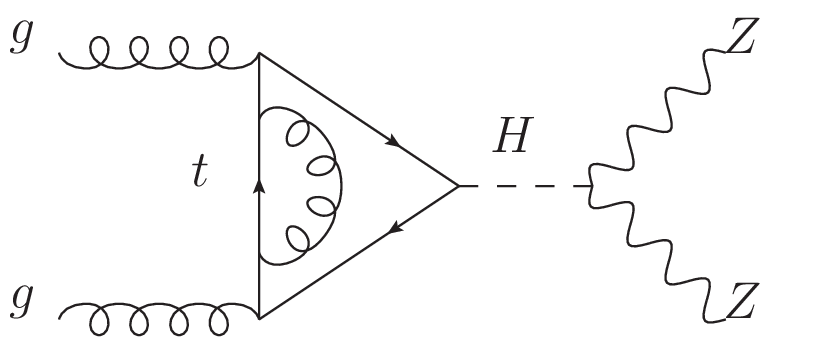}
  \caption{}
\end{subfigure}%
\begin{subfigure}{.3\textwidth}
  \centering
   \includegraphics[width=1.\linewidth]{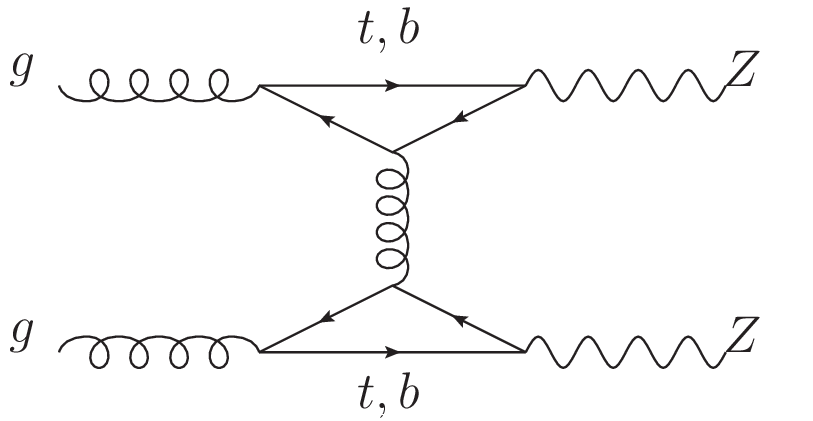}
  \caption{}
\end{subfigure} \\
\begin{subfigure}{.3\textwidth}
  \centering
     \includegraphics[width=1.\linewidth]{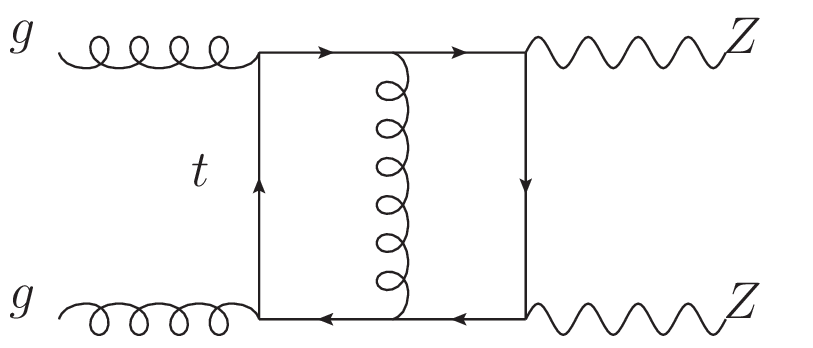}
  \caption{}
\end{subfigure}%
\begin{subfigure}{.3\textwidth}
  \centering
   \includegraphics[width=1.\linewidth]{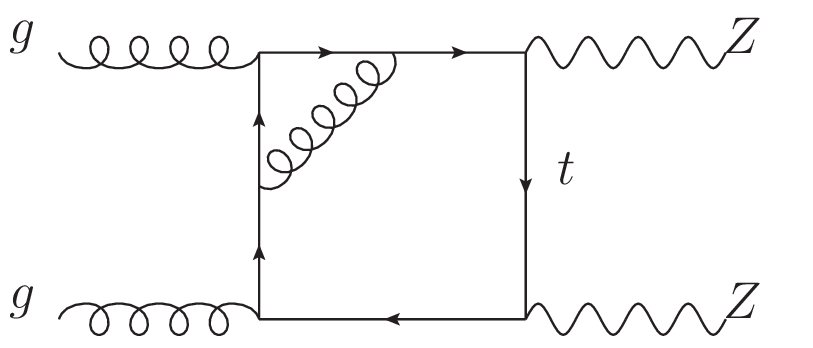}
  \caption{}
\end{subfigure}%
\begin{subfigure}{.3\textwidth}
  \centering
   \includegraphics[width=1.\linewidth]{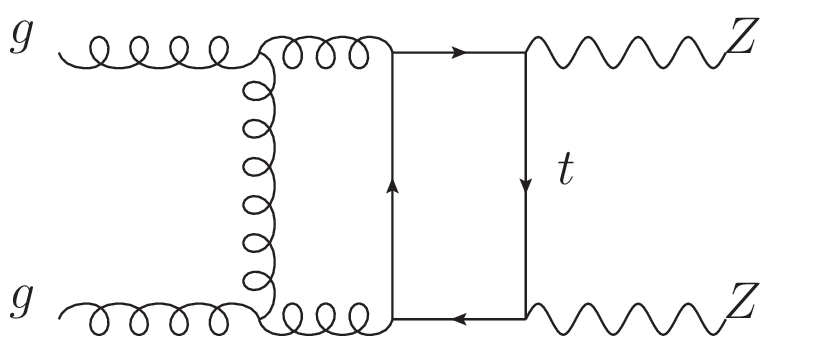}
  \caption{}
\end{subfigure} \\
\begin{subfigure}{.3\textwidth}
  \centering
     \includegraphics[width=1.\linewidth]{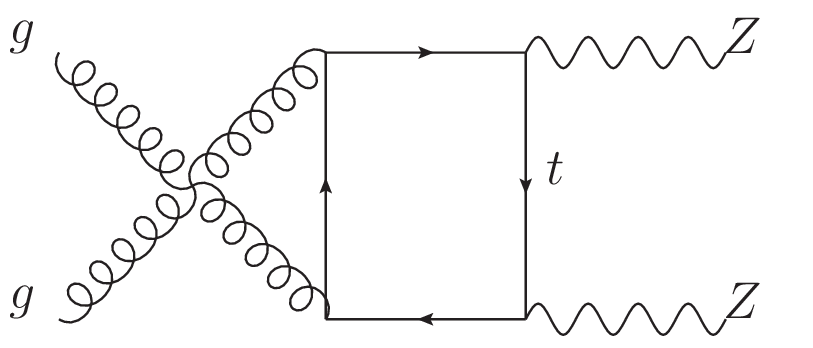}
  \caption{}
\end{subfigure}%
\begin{subfigure}{.3\textwidth}
  \centering
   \includegraphics[width=1.\linewidth]{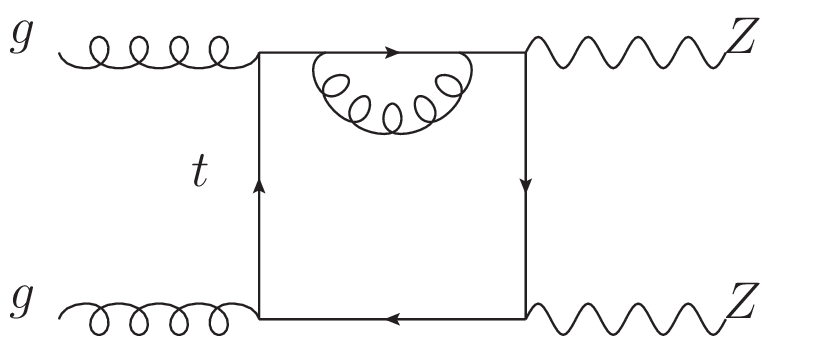}
  \caption{}
\end{subfigure}%

\caption{Representative Feynman diagrams contributing to the
  $gg \rightarrow ZZ$ amplitude at NLO. Loops of bottom quarks are included
  only in the double-triangle diagrams (c).}
\label{fig:NLOzz}
\end{figure}

We  consider a perturbative expansion of the form factors in the strong coupling
\begin{equation}
  \mathcal{A}_i = \mathcal{A}_i^{(0)} + \frac{\alpha_s}{\pi} \mathcal{A}_i^{(1)} +
  \mathcal{O}(\alpha_s^2)
\end{equation}
where one- and two-loop diagrams contribute respectively to the LO
$( \mathcal{A}_i^{(0)})$ and NLO $( \mathcal{A}_i^{(1)} )$. According to the
topology of the relevant Feynman diagrams (see fig.~\ref{fig:ggzzLO}), we identify a triangle and a box contribution to the LO form factors
\begin{equation}
\mathcal{A}_i^{(0)} = \mathcal{A}_i^{(0,\triangle)} + \mathcal{A}_i^{(0, \square)}.
\end{equation}
The above classification is modified at NLO, where the two-loop
triangle and box topologies (see fig.~\ref{fig:NLOzz}) are supplemented with
one-particle-reducible double-triangle diagrams as in
fig.~\ref{fig:NLOzz}(c). Therefore the NLO form factors are defined as 
\begin{equation}
  \mathcal{A}_i^{(1)} = \mathcal{A}_i^{(1, \triangle)} +
  \mathcal{A}_i^{(1, \square)} + \mathcal{A}_i^{(1, \bowtie)}.
\label{eq:nlosplitzz}
\end{equation}

The main result of this paper is the evaluation of the
$\mathcal{A}_i^{(1, \square)}$ using the $\pt$ expansion, which is
described below. For completeness, we provide also the results for
$\mathcal{A}_i^{(1, \triangle)}$ and $\mathcal{A}_i^{(1,\bowtie)}$.
We notice that only for $\mathcal{A}_i^{(1, \bowtie)}$ we include the
contributions from loops of bottom as well as top quarks, since the whole
fermion generation must be considered to remove the axial anomaly. 


With the previous definitions, the partonic cross section at LO is expressed as
\begin{equation}
\hat{\sigma}^{(0)} (\hat{s}) = \frac{\mz^4 G_F^2}{512 \pi \hat{s}^2} \left(\frac{\alpha_s (\mu_R)}{\pi}\right)^2 \int_{\hat{t}^-}^{\hat{t}^+} d \hat{t} \sum_i \left|\mathcal{A}_i^{(0)} \right|^2
\end{equation}
with $\hat{t}^\pm = 1/2 \left[-\hat{s} +2 \mz^2 \pm \sqrt{\hat{s}^2- \hat{s} ~ 4\mz^2 }\right] $.

\subsection{ The $\pt$-expansion method }
We now briefly recall the main points of the $\pt$ expansion.
The details of the method are presented in
refs.~\cite{Bonciani:2018omm,Alasfar:2021ppe}.
The Feynman amplitude is expanded in the transverse momentum at the integrand level via
the introduction of  the vector $r^\mu = p_1^\mu +p_3^\mu$, which satisfies the
relations
\begin{equation}
  r^2 = \hat{t} \qquad r \cdot p_1 = t' \qquad r \cdot p_2 = - t'
  \label{tpr}
\end{equation}
and can be written as
\begin{equation}
  r^\mu = \frac{t'}{s'}(p_2-p_1)^\mu +r_\perp^\mu,
  \label{spr}
\end{equation}
where the space-like vector $r_\perp^\mu$ is such that
\begin{equation}
r_\perp \cdot p_1 = r_\perp \cdot p_2 = 0,  \qquad r_\perp^2 = - \pt^2,
\end{equation} 
with
\begin{equation}
\pt^2 = \frac{\hat{t} \hat{u}-\mz^4}{\hat{s}}.
\end{equation}

In eqs.~(\ref{tpr},\ref{spr})  we introduce the primed Mandelstam variables 
\begin{equation}
  s' = p_1 \cdot p_2 = \frac{\hat{s}}{2}, \quad t' = p_1 \cdot p_3 =
  \frac{\hat{t}-\mz^2}{2}, \quad u' = p_2 \cdot p_3 =
  \frac{\hat{u}-\mz^2}{2},
\end{equation}
such that 
\begin{equation}
s' + t' +u' = 0, \qquad \pt^2 = \frac{2 t' u'}{s'} - \mz^2.
\end{equation}
The above relations allow to rewrite $t'$ as 
\begin{equation}
  t' = - \frac{s'}{2} \left \{ 1 \pm \sqrt{1 - 2 \frac{\pt^2 + \mz^2}{s'}}
  \right\}.
\label{eq:tprime}
\end{equation}

We are interested in the expansion of the amplitude in the forward
limit, corresponding to $t' \sim 0$, i.e~the minus-sign case in
eq.~\eqref{eq:tprime}. As discussed in refs.~\cite{Bonciani:2018omm,
  Alasfar:2021ppe}, the forward expansion is sufficient to obtain the
correct result for the cross section, if the (anti-)symmetry of the
form factors with respect to the exchange $t' \leftrightarrow u'$ is
ensured. The region of validity of the expansion is given by the condition
$\pt^2/(4\, \mt^2) \lesssim 1$.

The $\pt$ expansion of the relevant Feynman diagrams returns a result in terms
of the ratios of small vs large quantities, $x / y$, where
$x \in \{\pt^2, \mz^2 \}$ and $y \in \{ \hat{s},\mt^2\}$. We notice that, in the
$\pt$ expansion, the transverse momentum and $\mz$ are treated on the same
footing, i.e. a term $\mathcal{O}(\pt^{2}/\mz^2)$ is assumed to be
  $\mathcal{O}(1)$, not to be $\mathcal{O}(\pt^{2})$. 
Therefore the expansion can be also considered in terms of the
quantities $\mz^2$ and of the combination $ \pt^2 + \mz^2$ that enters in $t'$
(see eq.~(\ref{eq:tprime})).

After the scalar form factors $\mathcal{A}_i$ are expanded in the small
parameters above, and after the scalar integrals are decomposed along
a basis of master integrals (MI) using Integration-by-Parts (IBP)
identities, the form factors can be written as the following series
\begin{equation}
  \mathcal{A}_i = \mathcal{N}(\pt^2,\mz^2) \sum_{N=0}^{\infty} \sum_{i+j=N} c_{ij}
  (\pt^2)^i (\mz^2)^j,
\end{equation}
where the $c_{ij}$ coefficients are linear combinations of the MIs
resulting from the IBP reduction, which in turn depend only on
$\hat{s}$ and $\mt^2$, while $\mathcal{N}(\pt^2,\mz^2)$ is an overall
normalization factor which may depend on $\pt^2$ and $\mz^2$.

\section{Validation of the $\pt$ expansion at LO}
\label{sec:lo}
\begin{figure}
\centering
\begin{subfigure}{0.45\textwidth}
\hspace*{-1.cm}
    \includegraphics[scale=0.61]{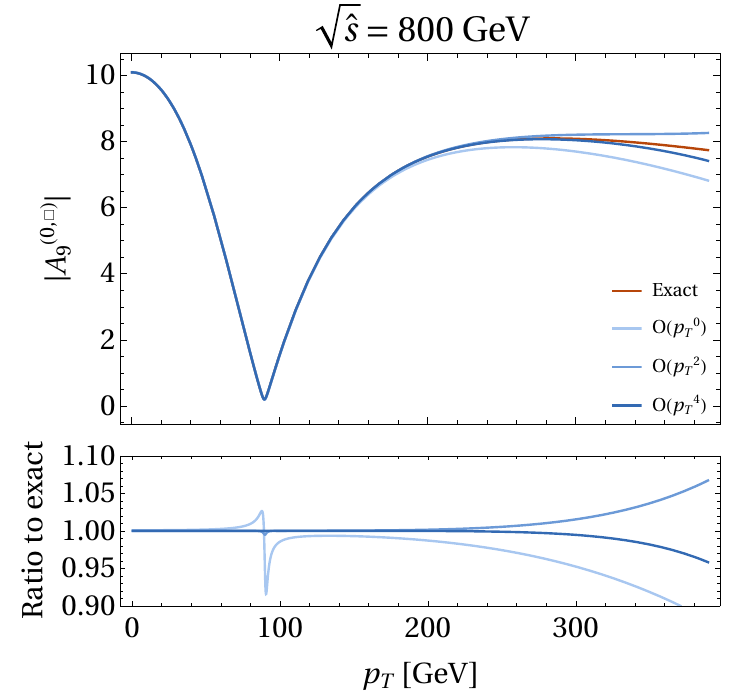}
  \caption{}
  \label{fig:ffloptonlya}
\end{subfigure} \quad
\begin{subfigure}{0.45\textwidth}
\hspace*{-0.8cm}
  \includegraphics[scale=0.61]{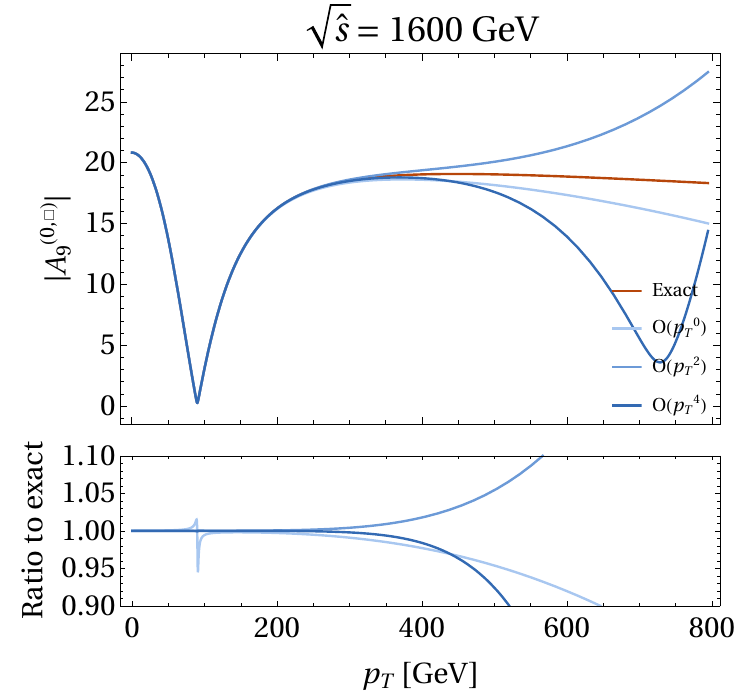}
  \caption{}
  \label{fig:ffloptonlyb}
\end{subfigure} \\
\caption{Absolute value of the form factor $\mathcal{A}_9^{(0, \square)}$  
  for moderate (a) and high (b) partonic centre-of-mass energies
  as a function of
  the transverse momentum. The exact result and the results obtained at
  various orders in the $\pt$ expansion are shown. 
}
\label{fig:fflopt}
\end{figure}
In this section we assess the level of accuracy of the $\pt$ expansion
in reproducing the known exact result at LO. We used
$\texttt{FeynArts}$~\cite{Hahn:2000kx} to generate the relevant
amplitudes, which we expanded in the limit of small $\pt$ using
private code. The latter relies on several functions implemented in
$\texttt{FeynCalc}$~\cite{Shtabovenko:2020gxv, Mertig:1990an}. The
expanded amplitude has been decomposed along a basis of MIs with
$\texttt{LiteRed}$~\cite{Lee:2013mka, Lee:2012cn}. 
The MIs can be expressed in terms of the $B_0$ and $C_0$ Passarino-Veltman
functions \cite{Passarino:1978jh} of argument $\mt^2$ and $\hat{s}$ or
$-\hat{s}$.  We also used
$\texttt{FeynCalc}$ to recompute the exact LO result for the amplitude
and found it in agreement\footnote{The comparison has been done for each form
factor, using the conversion formulas listed in
eqs.~\eqref{eq:convers_eqs_init}-\eqref{eq:convers_eqs_end} in
appendix~\ref{app:projezz}.} with the one provided in
ref.~\cite{Davies:2020lpf}.  Since at NLO only the box contributions will be
computed in an approximate way via the $\pt$ expansion,
in the following we focus on  the results of
the box form factors, $\mathcal{A}_i^{(0, \square)}$.
In particular, we consider the absolute value of the form
factor $\mathcal{A}_9^{(0, \square)}$ which provides the largest
contribution to the cross section. We notice that the 
$\mathcal{P}_9^{\mu \nu \rho \sigma}$ projector, as well as others,
exhibits a $\mathcal{O}(\pt^{-2})$ term (see eq.~(\ref{pnove}))
that can  question the fact that the corresponding form factor, 
$\mathcal{A}_9^{(0, \square)}$, must be regular in the limit $\pt \to 0$.
However,  the $\mathcal{O}(\pt^{-2})$ terms that appear in the projectors are
always multiplied by combination of  the
$S_{i}^{\mu \nu \rho \sigma} \: (i=17,\,18,\,19,\, 20)$ Lorentz structures that,
once the gauge condition eq.~(\ref{eq:gauge}) is enforced, exhibit an
$r_\perp^\mu\, r_\perp^\nu$ dependence. This  dependence ensures that
the form factors are regular in the limit $\pt \to 0$.

In fig.~\ref{fig:fflopt} we plot the exact result for the 
$\mathcal{A}_9^{(0, \square)}$ form factor, compared to various $\pt$-expanded
results for two values of the partonic centre-of-mass energy. 
The expanded-over-exact ratio in the bottom parts of the figure shows that the convergence
of the expansion is rather fast  and that already the
$\mathcal{O}(\pt^{2})$ result can reproduce the exact one with an
accuracy below the percent level. The spikes that are visible in the
expanded-over-exact ratio are due to the fact that the form factor is
zero near $\pt\sim 100$ GeV, and the ratio is not numerically
stable. From a comparison between figs.~\ref{fig:fflopt}(a) and
\ref{fig:fflopt}(b) we observe that, independently on the value of
$\sqrt{\hat{s}}$ the expansion is convergent only for values
$\pt \lesssim 300$ GeV. This is consistent with the limit
$\pt^2 \lesssim 4\, m_t^2$ that is assumed in the expansion.

\begin{table}
\centering
\begin{tabular}{ccccc}
\hline
$\mzz (\text{GeV})$ &  $\hat{\sigma}^{(0)} (\text{fb})$ \\
\hline
  & $\mathcal{O}(\pt^0)$ & $\mathcal{O}(\pt^2)$ & $\mathcal{O}(\pt^4)$ & Exact  \\
\hline
250 & 0.12486 & 0.12418 & 0.12412 & 0.12412\\
345 & 0.26619 & 0.26460 & 0.26448 & 0.26447\\
475 & 0.44887 & 0.44387 & 0.44368 & 0.44367\\
659 & 0.43135 & 0.38051 & 0.38644 & 0.38509\\
808 & 0.44108 & 0.30622 & 0.34712 & 0.32024\\
1255 & 0.70815 & 1.0303 & 0.66139 & 0.1883\\
\hline
\end{tabular}
\caption{The partonic LO cross section as a function of $\mzz$.
The exact result and the ones obtained at various orders
  in the $\pt$ expansion are shown. }
\label{tab:losig}
\end{table}

In table \ref{tab:losig} we present the values of the LO partonic cross section
for $gg \to ZZ$ as a function of the invariant mass of the two $Z$ bosons
system. The exact result is compared to various orders in the
$\pt$ expansion. The triangle contribution (see fig.~\ref{fig:ggzzLO}(a))
is evaluated always exactly.
 The value of the strong coupling is fixed at  $\alpha_s=0.118$, while we use as other parameters $ G_F = 1.1663786 \cdot 10^{-5}~\text{GeV}^{-2}, m_H = 125.1~\text{GeV}, m_Z = 91.1876~\text{GeV}, m_t = 173.21~\text{GeV}$.
The table shows that for $\mzz  \lesssim 500$ GeV already the
$\mathcal{O}(\pt^{2})$ term is in agreement with the exact result at the level
of less than 1 per mille. Increasing $\mzz$ up to 650 GeV the agreement
between the exact and the $\pt$-expanded result starts
to deteriorate reaching a  $\sim 1$ per cent difference. However,
the inclusion of the $\mathcal{O}(\pt^{4})$ term brings back
the difference to a few per mille level. For higher values of $\mzz$ the
most important contributions to the cross section come from regions in the
phase space where the $\pt$ expansion is not valid and the agreement
between the exact and the $\pt$-expanded results is  poor.

\section{Virtual corrections at NLO}
\label{sec:nlo}

Having showed that the $\pt$ expansion can provide accurate results
for the LO contribution, we move now to discuss its application to the
NLO QCD corrections, see fig.~\ref{fig:NLOzz} for representative Feynman diagrams.  In fact, we are going
to use the $\pt$ expansion only for the evaluation of the two-loop box
diagrams. For the two-loop one-particle-irreducible triangles we have
adapted the full analytic results of ref.~\cite{Aglietti:2006tp},
while  the reducible double triangles have been computed exactly
using $\texttt{FeynCalc}$ and checked with the results of ref.~\cite{Campbell:2016ivq}. With reference to
eq.~\eqref{eq:nlosplitzz}, we list the results for the form factors $
\mathcal{A}_i^{(1, \triangle)}$ and $\mathcal{A}_i^{(1, \bowtie)}$ in
appendix~\ref{app:analytzz}.

\subsection{Box diagrams}
\label{subsec:nlobox}

The evaluation of the two-loop box diagrams was carried out as follows.
 We generated the diagrams with
$\texttt{FeynArts}$ and performed all the manipulations of the
 amplitude within $\texttt{Mathematica}$ and $\texttt{FeynCalc}$.  The amplitude
 was  contracted with the 18 projectors in appendix  \ref{app:projezz}
and each form factor expanded 
 up to $\mathcal{O}(\pt^4)$. This led to the identification of a set of 9
families of scalar integrals. These families were analyzed using
$\texttt{LiteRed}$ and reduced to a basis of 52 known MIs
\cite{Aglietti:2006tp,Bonciani:2003te,Anastasiou:2006hc,Becchetti:2017abb,Caron-Huot:2014lda,vonManteuffel:2017hms},
which we found to be the same as those encoutered in $HH$ and $ZH$
production. Among the 52 MI, fifty can be expressed in terms of (generalised)
harmonic polylogarithms while two are elliptic integrals.
The evaluation of the (generalised) harmonic polylogarithms was done using
the code \texttt{handyG} \cite{Naterop:2019xaf}, while
the elliptic integrals were evaluated using the routines of
ref.~\cite{Bonciani:2018uvv}. The top quark mass was renormalised in the
on-shell scheme and the infra-red (IR) poles were subtracted as in
refs.~\cite{Degrassi:2016vss,Grober:2017uho}.

As a first check of our computation, we compared our $\pt$-expanded result
with the LME one presented in ref.~\cite{Davies:2020lpf}. It should be
noticed that $\pt$-expanded result actually “contains” the LME one. The LME
differs from the expansion in $\pt$ by the fact that $\hat{s}$ is treated as a
small parameter with respect to $\mt^2$ , and not on the same footing
as in the latter case. This implies that when the $\pt$-expanded result is
further expanded in terms of the $\hat{s}/\mt^2$ ratio the LME result has to be
recovered. This way, we were able to reproduce, at the analytic level, the
LME result of ref.~\cite{Davies:2020lpf} up to the order of our calculation.

\begin{table}
\centering
\begin{tabular}{ccc}
\hline
$\lambda_1 \lambda_2 \lambda_3 \lambda_4$  & Numerical & $\mathcal{O}(\pt^4)$  \\ 
\hline
$++++$ & 3.15549(8) + i\, 0.47235(8) & 3.16038 + i\,0.46980 \\ 
$+++-$ & 0.15950(7) + i\,0.14052(8) & 0.16305 + i\,0.13885 \\ 
$+-+-$ & -0.38609(7) + i\,0.10539(7) & -0.38617 + i\,0.11085 \\ 
$-+++$ & -0.46990(8) + i\,0.40207(8) & -0.46956 + i\,0.40506 \\ 
$+++0$ & 1.1248(2) - i\,0.0805(2) & 1.1256 - i\,0.0811 \\ 
$+-+0$ & -1.4803(2) + i\,0.4940(2) & -1.4799 + i\,0.4977 \\ 
$++00$ & 17.2585(6) + i\,29.5669(6) & 17.2602 + i\,29.5643 \\ 
$+-00$ & 10.2869(5) - i\,1.0571(6) & 10.2840 - i\,1.0640 \\ 
\hline 
\end{tabular}
\caption{Comparison of the NLO helicity amplitudes  between  the exact
  numerical values
  and the values given by  the $\pt$ expansion at $\mathcal{O}(\pt^4)$ in
  the kinematic point  provided in ref.~\cite{Agarwal:2020dye}.}
\label{tab:nlohel}
\end{table}

As a second check of our computation  we compare, in table~\ref{tab:nlohel},
the contribution of the two-loop
box diagrams at $\mathcal{O}(\pt^4)$ with the results of
ref.~\cite{Agarwal:2020dye}, which rely on the numerical evaluation of
the scalar integrals. The comparison is done at the level of the
helicity amplitudes\footnote{The helicities of the polarization vectors are
defined as:
$\epsilon_{\mu}^{\lambda_{1}} (p_1)\, \epsilon_{\nu}^{\lambda_{2}} (p_2)\,
\epsilon^{*\,\lambda_{3}}_\rho (p_3)\, \epsilon^{*\,\lambda_{4}}_\sigma (p_4)$.}
for the fixed phase-space point presented in
ref.~\cite{Agarwal:2020dye}, defined as $\parts = 142 /17 ~\mt^2$ and
$\partt= -125/22~\mt^2 \sim -5.7 ~\mt^2$. We remark that
the  IR subtraction scheme employed in ref.~\cite{Agarwal:2020dye}
($q_T$ subtraction)  differs from ours, so that to reproduce the form factors
of ref.~\cite{Agarwal:2020dye}, $\mathcal{A}_i^{(1) q_T}$,  
a shift in our form factors has to be introduced as follows:
\begin{equation}
\mathcal{A}_i^{(1) q_T} = \mathcal{A}_i^{(1)} + \frac{C_A}{4} \pi^2~.
\end{equation}
We notice that the point chosen  in ref.~\cite{Agarwal:2020dye} lies
somewhat beyond the region of validity of the $\pt$ expansion. Yet,
table \ref{tab:nlohel} shows that the relative difference between 
the  $\mathcal{O}(\pt^4)$ values and the numerical ones is, in general with
the exception of only one helicity  amplitude, at the per mille level and often
better, as in
the case of the amplitudes involving two longitudinally-polarized $Z$
bosons. The latter amplitudes are the dominant ones, as consequence of
the Goldstone Boson Equivalence Theorem.

\subsection{Merging the $\pt$ and High-Energy expansions}
\label{sec:merging}

The results we have presented so far allow to efficiently approximate
the two-loop box integrals with full top-mass dependence over a
specific region of the phase space, which spans the $ZZ$ production
threshold up to moderate partonic centre-of-mass energies (for the LHC). In
the high-energy region the $\pt$ expansion is still
accurate for any values of $\hat{s}$ and $|\hat{t}| \lesssim \,4
\mt^2$ or $|\hat{u}| \lesssim \,4 \mt^2$ but 
it cannot cover the complementary region $|\hat{t}|,|\hat{u}| \gtrsim 4 \mt^2$ that becomes allowed by the kinematics.

In ref.~\cite{Bellafronte:2022jmo} we showed that the $\pt$ expansion
and the HE expansion are accurate in complementary
phase-space regions. The HE expansion is an expansion in terms of the
ratios $x/y$ where $x \in \{\mt^2, \mz^2 \}$ and $y \in \{ \hat{s}, \hat{t}, \hat{u} \}$
and it is valid in phase-space regions where both $| \hat{t}|/(4 \,\mt^2) \gtrsim 1$ and $| \hat{u}|/(4 \,\mt^2) \gtrsim 1$.
We also showed that if each expansion is extended beyond
its border of validity
via the use of Padé approximants the two results can be merged into a
single prediction that is accurate over the complete phase space.

Results for the evaluation of the two-loop box contribution in the $gg \to ZZ$
process via the HE expansion were
presented in ref.~\cite{Davies:2020lpf}.
Applying  the method of ref.~\cite{Bellafronte:2022jmo}
we merge our results for the box contribution 
with those obtained in  ref.~\cite{Davies:2020lpf}. Below we recall the
main steps of the merging method:
\begin{enumerate}
\item We use the formulas in
  eqs.~\eqref{eq:convers_eqs_init}-\eqref{eq:convers_eqs_end} to adapt
  the form factors of ref.~\cite{Davies:2020lpf} to our decomposition
  of the amplitude.
\item We rewrite the small quantities on which the $\pt$ and the HE
  expansions are based in terms of a scaling parameter $x$.
\item We construct Padé approximants for each form factor, defined as
\begin{equation}
[m/n](x) = \frac{p_0 + p_1 x + \dots + p_m x^m}{1+ q_1 x + \dots  q_n x^n},
\label{eq:mnpade}
\end{equation}
where $p_i, q_i$ are expressed as linear combinations of the
coefficients of each expansion. We refer to these approximants as the
$\pt$-Padé and HE-Padé.
\item In evaluating the amplitude, we use the HE-Padé approximant for
  phase-space points such that $|\hat{t}| > 4\mt^2$ and $|\hat{u}| >
  4\mt^2$ whereas for all the remaining phase-space points we use the
  $\pt$-Padé approximant.
\end{enumerate}

With our three terms in the $\pt$ expansion we constructed a [1/1] 
$\pt$-Padé. The several terms in the HE expansion presented in
ref.~\cite{Davies:2020lpf} allow to construct different [x/y] HE-Padé. 
We construct both the [5/5] and [6/6] HE-Padé and compare the results
using the different orders. As a result we found that both the [5/5] and
[6/6] HE-Padé give very similar results.\\

\section{Results}
\label{sec:results}
\begin{figure}
\centering
\begin{subfigure}{0.45\textwidth}
\hspace*{-0.8cm}
    \includegraphics[scale=0.52]{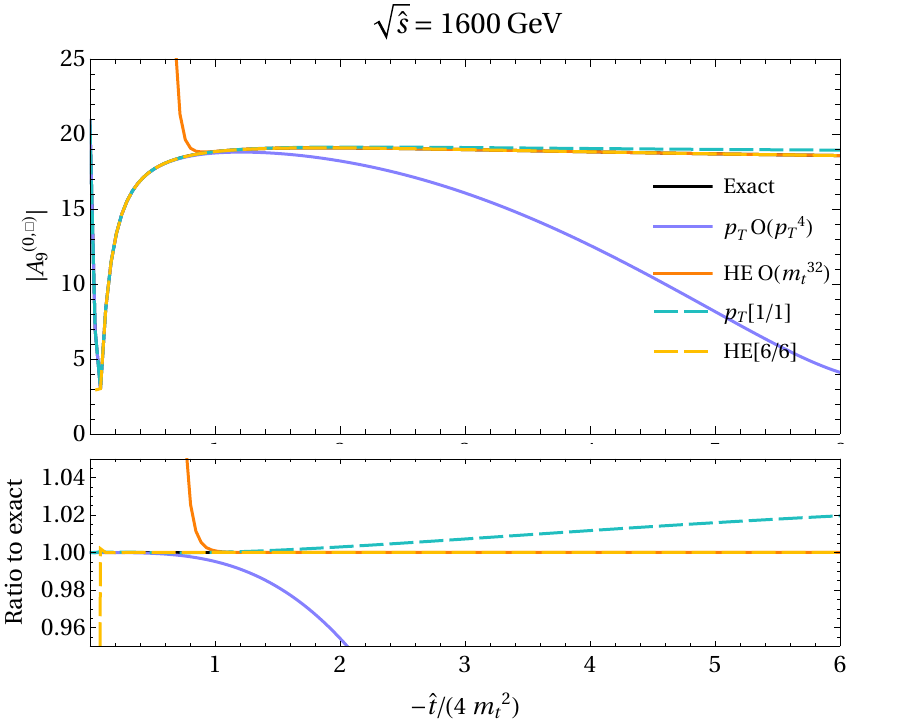}
  \caption{}
  \label{fig:fflomerged}
\end{subfigure} \quad
\begin{subfigure}{0.45\textwidth}
\vspace*{0.3cm}
\hspace*{-0.4cm}
  \includegraphics[scale=0.52]{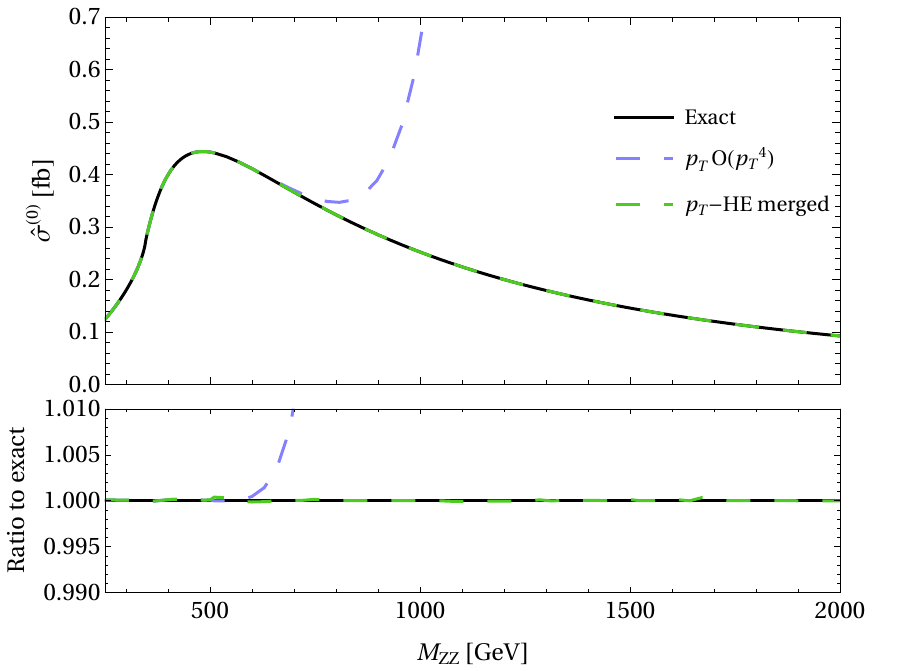}
  \caption{}
  \label{fig:xseclomerged}
\end{subfigure} \\
\caption{(a) Absolute value of a form factor for a fixed
  $\sqrt{\hat{s}}$ using the $\pt$ and the HE expansion and the best
  Padé approximants. (b) Partonic cross section at LO using the exact
  result, the $\pt$ expansion and the prediction using the merged
  $\pt$ and HE Padé approximants. }
  \label{fig:lomergedvalid}
\end{figure}

Before applying the merging approach for the form factors at NLO, we
verified its reliability testing it against  the exact LO
contribution. In fig.~\ref{fig:lomergedvalid}(a)  the exact result (solid black line)
for the form factor, $\mathcal{A}_9^{(0, \square)}$ as a function of
$- \hat{t}/(4 \,\mt^2)$  is compared to different
approximate evaluations. In the figure the fixed order results
$\mathcal{O}(\pt^{4})$ in the $\pt$ expansion (solid blue) and
$\mathcal{O}(\mt^{32})$ in the HE expansion (solid orange) are shown
against the Padé-improved versions, [1/1] $\pt$-Padé (dashed light blue)
and  $[6/6]$ HE-Padé (dashed yellow).  The bottom panel
shows the ratio of the different approximations over the exact
result. It is evident that the region $|\hat{t}| \sim 4\, \mt^2$, where
both the $\pt$ and HE expansions (solid blue and orange) begin to
diverge, is well covered by the respective Padé approximants, so that
one has a way to accurately reproduce the exact result for any value
of $|\hat{t}|$. In fig.~\ref{fig:lomergedvalid}(b) we compare different
predictions at the level of the LO partonic cross section. One can see
that the accuracy of the merged result (dashed green) is at the level
of per mille or below, an order of magnitude better than what observed
in ref.~\cite{Bellafronte:2022jmo}. We notice that, this remarkable
degree of accuracy is not achieved for each individual form factor. However, 
the impact of larger deviations is suppressed with the integration
over $\hat{t}$.


\begin{figure}[t]
\centering
  \includegraphics[scale=0.8]{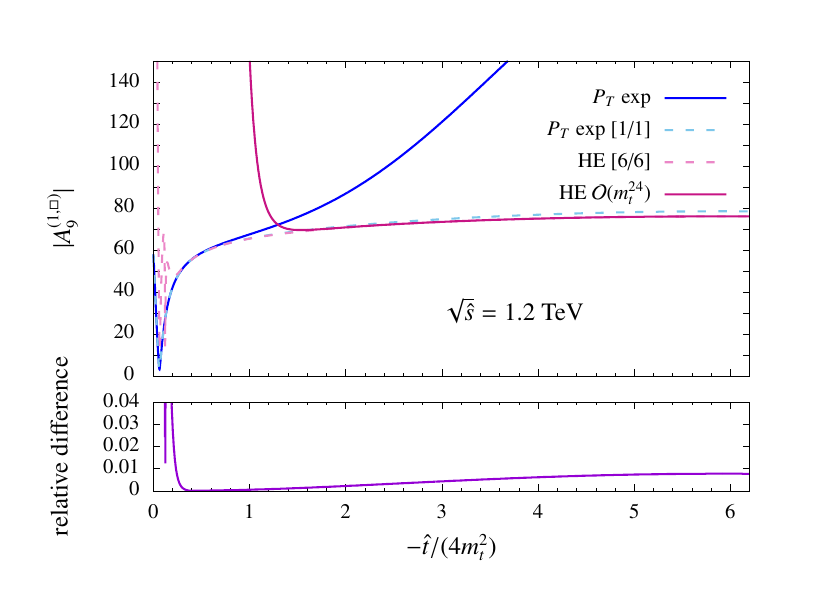}
  \caption{NLO form factor for $\sqrt{\hat{s}}=1200\text{ GeV}$ in
    various approximations: $\pt$ expansion (solid blue), [1/1] Padé
    approximant based on the $\pt$ expansion (dashed, light blue),
    high-energy expansion (solid, pink) and [6/6] HE Padé
    approximant (dashed, rosa). The lower panel shows the relative
    difference $\Delta$ (see text).}
\label{fig:ffnlomerged}
\end{figure}

\begin{figure}
\centering
\begin{subfigure}{1.\textwidth}
\hspace*{-1.8cm}
    \includegraphics[scale=0.75]{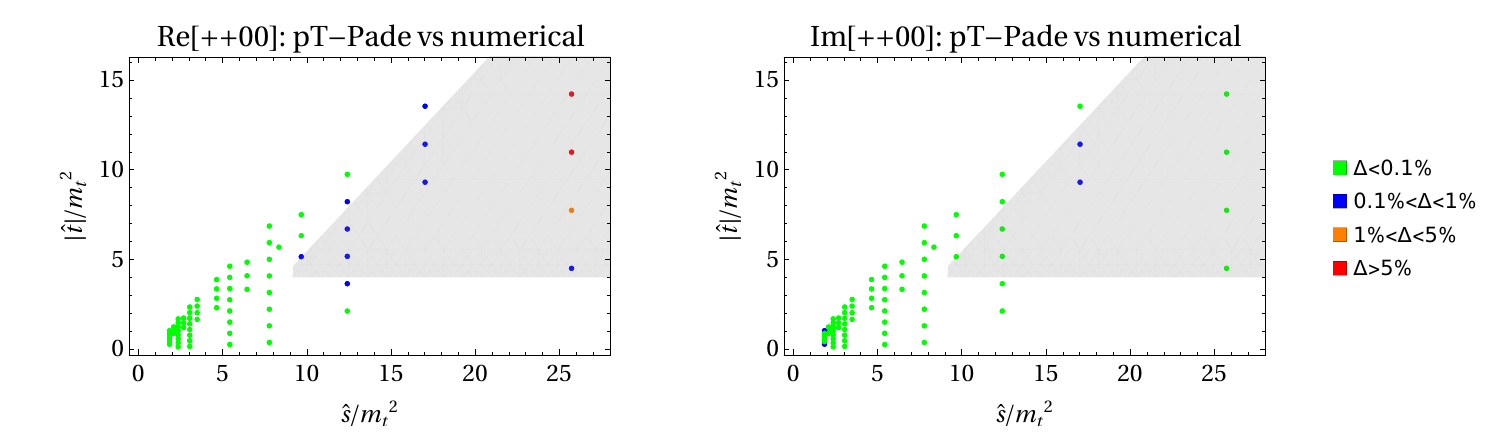}
  \caption{}
  \label{fig:repadept}
\end{subfigure} \quad

\begin{subfigure}{1.\textwidth}
\hspace*{-1.8cm}
  \includegraphics[scale=0.75]{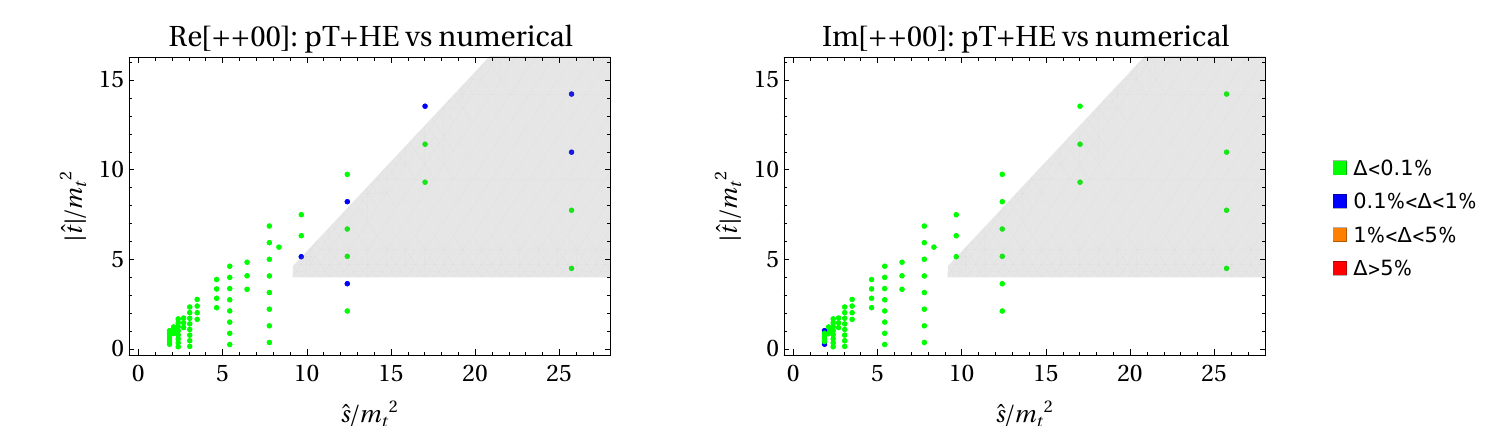}
  \caption{}
  \label{fig:impadept}
\end{subfigure} \\
\caption{Relative difference between several phase-space points of ref.~\cite{Agarwal:2020dye} and different approximations for the helicity amplitude ++00. Points in the shaded region are outside the formal limit of validity of the $\pt$ expansion. (a) [1/1] $p_T$-Padé only; (b) merging of the $p_T$-Padé and the HE-Padé.}
\label{fig:ptpade}
\end{figure}

We now present our results at NLO. First,  we checked their the validity in
two ways:
\begin{itemize}
\item We compared the $\pt$-Padé and HE-Padé in the vicinity of
  $|\hat{t}| \sim 4 m_t^2$, for several values of $\hat{s}$ in the
  range [800, 2000] GeV, ensuring that the two analytic approximations
  are indeed complementary and that the matching point is chosen
  well. In fig.~\ref{fig:ffnlomerged}  the NLO
  form factor $\mathcal{A}_9^{(1, \square)}$ is shown as a function of
  $-\hat{t}/( 4\, \mt^2)$ for both fixed-order $\pt$ and HE expansions
  and Padé-improved results. The lower panel of the figure
  shows the relative difference, $\Delta$, between the $\pt$-Padé
   and the HE-Padé, defined as
\be  
  \Delta =2
    \left|\frac{[1/1]_{\pt}-[6/6]_\text{HE}}{[1/1]_{\pt}+[6/6]_\text{HE}} \right|,
\ee
  where one can see that
  at $|\hat{t}| \sim 4 m_t^2$ the two expansion are in good agreement (on
  the level of 0.5 per mille).

\item We compared our merged results with the helicity amplitudes of
  ref.~\cite{Agarwal:2020dye}, obtained from a numerical evaluation of
  the scalar integrals. The results for several phase-space points
  were provided privately by the authors of
  ref.~\cite{Agarwal:2020dye}. In fig.~\ref{fig:ptpade}
  we plot the results for the amplitude $++00$  using different
  colors depending on the level of agreement $\Delta$ between the numerical and
  merged prediction. The shaded region in the plots corresponds to the high-energy region, where $|\hat{t}| > 4\mt^2$ and $|\hat{u}| >
  4\mt^2$. In fig.~\ref{fig:ptpade}~(a) we compare the numerical evaluation with our results using only the [1/1] $\pt$-Padé. Whereas the agreement is generally better than per mille in the region of validity of the $\pt$ expansion, deviations larger than 5\% are visible in the high-energy region for the real part. When we consider the merged prediction, the agreement with the numerical evaluation is improved, and the majority of the phase space points have differences below the per mille level.
  It must be noted that, as seen in table~\ref{tab:nlohel}, this level of agreement is not shared among all the helicity amplitudes. This is however a consequence of the relations used to convert our results in terms of form factors into helicity amplitudes. Indeed, for some phase-space points, we observed delicate cancellations in the combination of the form factors. At the same time, for these particular phase-space points, the corresponding contributions to the helicity amplitudes are tiny. Therefore, we expect an overall accuarcy at the subpercent level for our results. 
  
\end{itemize}

\begin{figure}
\centering
  \includegraphics[scale=0.8]{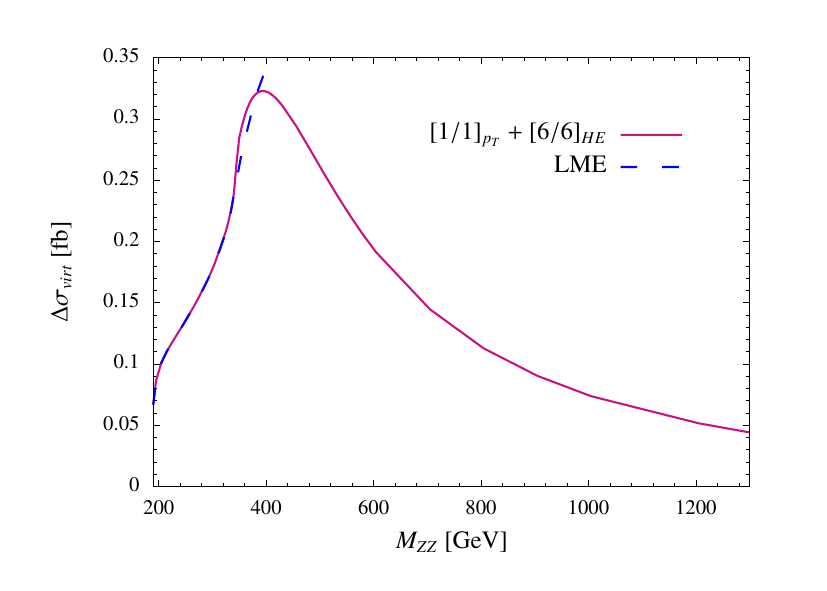}
  \caption{The partonic virtual corrections as a function of $\mzz$ merging the $\pt$ expansion and the HE (pink solid line) compared to the virtual corrections obtained when using the large mass expanded results for the NLO box and triangles (dashed blue line). }
  \label{fig:nlovirtxsec}
\end{figure}

In fig.~\ref{fig:nlovirtxsec} one can see our result for the finite part of the virtual
corrections defined as
\begin{equation}
\Delta \sigma_\text{virt} = \int_{\hat{t}^-}^{\hat{t}^+} d \hat{t} \frac{1}{2} \frac{1}{16 \pi \hat{s}^2} \left( \frac{\alpha_s (\mu_R)}{\pi} \right) \mathcal{V}_\text{fin} (\hat{t})
\label{eq:sigmavirt}
\end{equation}
where the finite part of the virtual correction is
\begin{equation}
\mathcal{V}_\text{fin} = \frac{G_F^2 \mz^4}{16} \left( \frac{\alpha_s (\mu_R)}{\pi} \right)^2  \left\{ \sum_{i} \left|\mathcal{A}_i^{(0)} \right|^2\frac{C_A}{2}\left(\pi^2- \log^2\left(\frac{\mu_R^2}{\hat{s}}\right)\right)  +2\sum_i\text{Re}\left[\mathcal{A}_i^{(0)}\left(\mathcal{A}_i^{(1)}\right)^* \right] \right\}.
\end{equation}
We used $\mu_R = \mzz/2$ as renormalisation scale.
 The pink line shows the merged result of
a [1/1] $\pt$-expanded Padé approximant and a [6/6] HE Padé
approximant. For comparison we show the large mass expanded results of
ref.~\cite{Davies:2020lpf}.  Within the range of validity of the large mass expansion the results
agree very well. We have checked that using a [5/5] HE Padé
approximant would change the result only below the integration error.


\section{Conclusion}
\label{sec:concl}
The accurate description of the process $gg \to ZZ$ is needed in order to improve the extraction of the total width of the Higgs boson, as well as a precision test for the SM. In this paper we have presented the calculation of the top-quark loops in the virtual QCD corrections at NLO. Our main result is the computation of the box diagrams using the $\pt$ expansion, which gives accurate results in a phase-space region that so far has not been covered by other analytic approximations. 

As a by-product of our calculation, we have used the results obtained in ref.~\cite{Davies:2020lpf} to complement our analytic calculation, showing that a combination of the two approaches can provide an efficient approximation for the cross section, at a level of accuracy that is more than adequate for phenomenological applications. We emphasise that our analytic approach is very flexible and can be included into Monte Carlo programmes, as was already done for the case of Higgs pair production, $gg\to HH$ \cite{Bagnaschi:2023rbx}.  It allows to change easily the input parameters, which is for instance necessary to evaluate the top-mass renormalisation scheme uncertainty that is expected to be large in the top-mediated contributions. Indeed, it is currently the  largest uncertainty for the $gg\to HH$ \cite{Baglio:2018lrj, Baglio:2020wgt} and $gg\to ZH$ \cite{Degrassi:2022mro} processes. In addition, the flexibility of the approach allows for a straightforward application to beyond-the-SM scenarios.

We recall that for a full description of $gg \to ZZ$ at NLO in QCD also the contribution mediated by loops of light quarks needs to be considered \cite{vonManteuffel:2015msa,Caola:2015psa}, as well as the one from real-emission diagrams \cite{Grazzini:2018owa}. We leave the combination of these effects with our results to future work.

\section*{Acknowledgments}
We thank Luigi Bellafronte and Pier Paolo Giardino for participating in the
early stages of this project. We are greatful to the authors of
ref.~\cite{Agarwal:2020dye} for providing us with the results for
additional phase-space points and for useful discussions.
The work of RG is supported in part by the Italian MUR
Departments of Excellence grant 2023-2027 ”Quantum Frontiers”
 and the ICSC – Centro Nazionale di Ricerca in High Performance Computing, 
 Big Data and Quantum Computing, funded by European Union – NextGenerationEU.  The
authors acknowledge support from the COMETA COST Action CA22130.  The
diagrams in this paper have been drawn using
$\texttt{JaxoDraw}$~\cite{Binosi:2003yf}.
\newpage
\begin{appendletterA}
\section{Projectors}
\label{app:projezz}
We present the explicit expressions of the orthonormal projectors  $\mathcal{P}_i^{\mu\nu\rho \sigma}$
appearing in eq.~(\ref{eq:clproje}). Following the notation of section~\ref{sec:def}, the antisymmetric projectors under the exchange $\{\mu \leftrightarrow \nu$ , $p_1 \leftrightarrow p_2\}$ are
\begin{align}
\mathcal{P}_1^{\mu \nu \rho \sigma}&= \frac{\mz}{\sqrt{2}\pt} \Biggl[\frac{1}{\pt^2 s'} \left(S_{17}^{\mu \nu \rho \sigma}-S_{20}^{\mu \nu \rho \sigma}\right) \Biggr]\\
\mathcal{P}_2^{\mu \nu \rho \sigma} &= \frac{\mz}{\sqrt{2}\pt} \Biggl[ \frac{1}{\pt^2 s'}\left( \frac{t' }{u'} S_{19}^{\mu \nu \rho \sigma}-\frac{u' }{t'} S_{18}^{\mu \nu \rho \sigma}\right) \Biggr] \\
\mathcal{P}_3^{\mu \nu \rho \sigma}&= \frac{\mz}{\sqrt{2}\pt} \Biggl[\frac{S_{5}^{\mu \nu \rho \sigma}}{u'}-\frac{S_{6}^{\mu \nu \rho \sigma}}{t'}-\frac{1}{\pt^2 s'} \left(S_{17}^{\mu \nu \rho \sigma}-S_{20}^{\mu \nu \rho \sigma}+\frac{u' }{t'}S_{18}^{\mu \nu \rho \sigma}-\frac{t' }{u'}S_{19}^{\mu \nu \rho \sigma}\right) \Biggr] \\
\mathcal{P}_4^{\mu \nu \rho \sigma}&= \frac{\mz}{\sqrt{2}\pt} \Biggl[ \frac{1}{s'} \left(S_{9}^{\mu \nu \rho \sigma}-S_{12}^{\mu \nu \rho \sigma}\right)+\frac{1}{\pt^2 s'} \left(S_{17}^{\mu \nu \rho \sigma}-S_{20}^{\mu \nu \rho \sigma}\right) \Biggr] \\
\mathcal{P}_5^{\mu \nu \rho \sigma}&= \frac{\mz}{\sqrt{2}\pt} \Biggl[ \frac{1}{s'} \left(\frac{t' 
}{ u'}S_{11}^{\mu \nu \rho \sigma}-\frac{u' }{ t'} S_{10}^{\mu \nu \rho \sigma}\right) +\frac{1}{\pt^2 s'} \left(\frac{t' }{u'} S_{19}^{\mu \nu \rho \sigma}-\frac{u'}{t'} S_{18}^{\mu \nu \rho \sigma}\right) \Biggr] \\
\mathcal{P}_6^{\mu \nu \rho \sigma}&= \frac{\mz}{\sqrt{2}\pt} \Biggl[ \frac{S_{13}^{\mu \nu \rho \sigma}}{u'}-\frac{S_{16}^{\mu \nu \rho \sigma}}{t'}-\frac{1}{\pt^2 s'} \left(S_{17}^{\mu \nu \rho \sigma}-S_{20}^{\mu \nu \rho \sigma}+\frac{t' }{u'} S_{19}^{\mu \nu \rho \sigma}-\frac{u'}{t'} S_{18}^{\mu \nu \rho \sigma}\right) \Biggr] \\
\mathcal{P}_7^{\mu \nu \rho \sigma}&= \frac{\mz^2}{\sqrt{2}\pt^2} \Biggl[\frac{\left(\mz^2-\pt^2\right) }{2 \mz^2} \left(\frac{S_{13}^{\mu \nu \rho \sigma}}{u'}-\frac{S_{16}^{\mu \nu \rho \sigma}}{t'}\right)+\frac{1}{\mz^2 s'}\left(u' S_{14}^{\mu \nu \rho \sigma}-t' S_{15}^{\mu \nu \rho \sigma}\right)  \\
&+\frac{1}{\mz^2 s'}\left(S_{17}^{\mu \nu \rho \sigma}-S_{20}^{\mu \nu \rho \sigma}+\frac{t' }{u'} S_{19}^{\mu \nu \rho \sigma}-\frac{u'}{t'} S_{18}^{\mu \nu \rho \sigma}\right) \nonumber \Biggr] \\
\mathcal{P}_8^{\mu \nu \rho \sigma}&= \frac{\mz^2}{\sqrt{2}\pt^2} \Biggl[ \frac{1}{\mz^2 s'}\left(u' S_{4}^{\mu \nu \rho \sigma}-t' S_{7}^{\mu \nu \rho \sigma}\right)+\frac{\left(\mz^2-\pt^2\right)}{2 \mz^2} \left(\frac{S_{5}^{\mu \nu \rho \sigma}}{u'}-\frac{S_{6}^{\mu \nu \rho \sigma}}{t'}\right) \\ 
&+\frac{1}{\mz^2 s'}\left(S_{17}^{\mu \nu \rho \sigma}-S_{20}^{\mu \nu \rho \sigma}+\frac{u'}{t'} S_{18}^{\mu \nu \rho \sigma}-\frac{t' }{u'} S_{19}^{\mu \nu \rho \sigma}\right) \nonumber \Biggr].
\end{align}

The symmetric projectors are
\begin{align}
  \mathcal{P}_9^{\mu \nu \rho \sigma}&= \frac{\mz^2}{\sqrt{\pt^4+\mz^4}} \Biggl[\frac{1}{\pt^2 s'} \left(S_{17}^{\mu \nu \rho \sigma}+S_{20}^{\mu \nu \rho \sigma}\right) \Biggr]
  \label{pnove} \\
\mathcal{P}_{10}^{\mu \nu \rho \sigma}&= \frac{\sqrt{\pt^4+\mz^4}}{2 \pt^2} \Biggl[\frac{\left(\mz^4-\pt^4\right) }{ \left(\mz^4+\pt^4\right) \pt^2 s'} \left(S_{17}^{\mu \nu \rho \sigma}+S_{20}^{\mu \nu \rho \sigma}\right) +\frac{1}{\pt^2 s'}\left(\frac{u'}{t'} S_{18}^{\mu \nu \rho \sigma}+\frac{t' }{u'} S_{19}^{\mu \nu \rho \sigma}\right) \Biggr] \\
\mathcal{P}_{11}^{\mu \nu \rho \sigma}&= \frac{\mz^2}{\sqrt{\pt^4+\mz^4}} \Biggl[ \frac{1}{s'} \left(S_{9}^{\mu \nu \rho \sigma}+S_{12}^{\mu \nu \rho \sigma}\right)+\frac{1}{\pt^2 s'} \left(S_{17}^{\mu \nu \rho \sigma}+S_{20}^{\mu \nu \rho \sigma}\right) \Biggr] \\
\mathcal{P}_{12}^{\mu \nu \rho \sigma}&= \frac{\sqrt{\pt^4+\mz^4}}{2 \pt^2} \Biggl[ \frac{\left(\mz^4-\pt^4\right)}{\left(\mz^4+\pt^4\right) s'}\left(S_{9}^{\mu \nu \rho \sigma}+S_{12}^{\mu \nu \rho \sigma}\right)+\frac{1}{s'}\left(\frac{u' \
}{ t'} S_{10}^{\mu \nu \rho \sigma} +\frac{t' }{ u'} S_{11}^{\mu \nu \rho \sigma}\right)   \\
&+\frac{\left(\mz^4-\pt^4\right) }{ \left(\mz^4+\pt^4\right) \pt^2 s'} \left(S_{17}^{\mu \nu \rho \sigma}+S_{20}^{\mu \nu \rho \sigma}\right) +\frac{1}{\pt^2 s'}\left(\frac{u'}{t'} S_{18}^{\mu \nu \rho \sigma}+\frac{t' }{u'} S_{19}^{\mu \nu \rho \sigma}\right) \nonumber \Biggr] \\
\mathcal{P}_{13}^{\mu \nu \rho \sigma} &= \frac{\mz}{\sqrt{2}\pt} \Biggl[ \frac{S_{13}^{\mu \nu \rho \sigma}}{u'}+\frac{S_{16}^{\mu \nu \rho \sigma}}{t'}-\frac{1}{\pt^2 s'} \left(S_{17}^{\mu \nu \rho \sigma}+S_{20}^{\mu \nu \rho \sigma}\right) -\frac{1}{\pt^2 s'}\left(\frac{u'}{t'} S_{18}^{\mu \nu \rho \sigma}+\frac{t' }{u'}   S_{19}^{\mu \nu \rho \sigma}\right) \Biggr] \\
\mathcal{P}_{14}^{\mu \nu \rho \sigma}&= \frac{\mz^2}{\sqrt{2}\pt^2} \Biggl[\frac{\left(\mz^2-\pt^2\right) }{2 \mz^2 }\left(\frac{S_{13}^{\mu \nu \rho \sigma}}{u'}+\frac{S_{16}^{\mu \nu \rho \sigma}}{t'}\right)+ \frac{1}{\mz^2 s'}\left(u' S_{14}^{\mu \nu \rho \sigma}+t' S_{15}^{\mu \nu \rho \sigma}\right)  \\
&-\frac{1}{\pt^2 s'} \left(S_{17}^{\mu \nu \rho \sigma}+S_{20}^{\mu \nu \rho \sigma}+\frac{u'}{t'} S_{18}^{\mu \nu \rho \sigma}+\frac{t' }{u'} S_{19}^{\mu \nu \rho \sigma}\right) \nonumber  \Biggr] \\
\mathcal{P}_{15}^{\mu \nu \rho \sigma}&= \frac{\sqrt{2} \mz^3}{\pt (\pt^2+\mz^2)} \Biggl[ \frac{1}{\mz^2 s'}\left(u' S_{4}^{\mu \nu \rho \sigma}+t' S_{7}^{\mu \nu \rho \sigma}\right)  \\
&+\frac{\left(\mz^2+\pt^2\right) }{2 \mz^2 \pt^2 s'} \left(S_{17}^{\mu \nu \rho \sigma}+S_{20}^{\mu \nu \rho \sigma}+\frac{u'}{t'} S_{18}^{\mu \nu \rho \sigma}+\frac{t' }{u'} S_{19}^{\mu \nu \rho \sigma}\right) \Biggr] \nonumber \\
\mathcal{P}_{16}^{\mu \nu \rho \sigma}&= \frac{\pt^2+\mz^2}{2 \sqrt{2} \pt^2} \Biggl[ \frac{2 \mz^2}{s' \pt^2 (\pt^2+\mz^2)}\left(S_{17}^{\mu \nu \rho \sigma}+S_{20}^{\mu \nu \rho \sigma}+\frac{u'}{t'} S_{18}^{\mu \nu \rho \sigma}+\frac{t' }{u'} S_{19}^{\mu \nu \rho \sigma}\right) \\
&-\frac{2 \left(\pt^2-\mz^2\right) }{\left(\mz^2+\pt^2\right)^2 s'} \left(u' S_{4}^{\mu \nu \rho \sigma}+t' S_{7}^{\mu \nu \rho \sigma}\right)+\frac{S_{5}^{\mu \nu \rho \sigma}}{u'}+\frac{S_{6}^{\mu \nu \rho \sigma}}{t'} \Biggr] \nonumber \\
\mathcal{P}_{17}^{\mu \nu \rho \sigma} &= \frac{\pt^2+2 \mz^2}{2 \mz^2} \Biggl\{ \frac{\mz^2}{2 \mz^2+\pt^2} \left(S_{2}^{\mu \nu \rho \sigma} +S_{3}^{\mu \nu \rho \sigma}\right)-\frac{\mz^2}{(\pt^2 + 2 \mz^2) s' \pt^2} \left(u' S_{4}^{\mu \nu \rho \sigma} + t' S_{7}^{\mu \nu \rho \sigma}\right) \\
&-\frac{\left(\mz^2+\pt^2\right) \mz^2}{\pt^2 \left(2 \mz^2+\pt^2\right)} \Biggl[\frac{S_{5}^{\mu \nu \rho \sigma}}{2 u'}+\frac{S_{6}^{\mu \nu \rho \sigma}}{2 t'}-\frac{S_{13}^{\mu \nu \rho \sigma}}{2 u'}-\frac{S_{16}^{\mu \nu \rho \sigma}}{2 t'}+\frac{1}{\pt^2 s'} \Bigl(S_{17}^{\mu \nu \rho \sigma}+S_{20}^{\mu \nu \rho \sigma}   \nonumber \\
&+\frac{u'}{t'} S_{18}^{\mu \nu \rho \sigma}+\frac{t' }{u'} S_{19}^{\mu \nu \rho \sigma}\Bigr) \Biggr] +\frac{ \mz^2}{ \pt^2 s' \left(2 \mz^2+\pt^2\right)}\left(u' S_{14}^{\mu \nu \rho \sigma}+t' S_{15}^{\mu \nu \rho \sigma}\right) \Biggr\} \nonumber 
\end{align}

\begin{align}
\mathcal{P}_{18}^{\mu \nu \rho \sigma}&=\frac{S_{8}^{\mu \nu \rho \sigma}}{\pt^2}-\frac{\left(\mz^2+\pt^2 \right)}{2 \pt^4 s'}\left(S_{17}^{\mu \nu \rho \sigma}+S_{20}^{\mu \nu \rho \sigma}\right)+\frac{u' \
\left(-\mz^4+\pt^4-2 \pt^2 (s'-t'+u')\right) S_{18}^{\mu \nu \rho \sigma}}{2 \pt^4 \
\left(\mz^2+\pt^2\right) s' t'} \\
& -\frac{t' \
\left(\mz^4-\pt^4+2 \pt^2 (s'+t'-u')\right) S_{19}^{\mu \nu \rho \sigma}}{2 \pt^4 \
\left(\mz^2+\pt^2\right) s' \
u'}. \nonumber
\end{align}
Finally, we include the last two projectors, which have null norm in $D=4$ dimensions and thus do not contribute to the amplitude.
\begin{align}
\mathcal{P}_{19}^{\mu \nu \rho \sigma}&=S_{1}^{\mu \nu \rho \sigma}-\frac{S_{2}^{\mu \nu \rho \sigma}}{2}-\frac{S_{3}^{\mu \nu \rho \sigma}}{2}+\frac{1}{2 \
\pt^2 s'}\left(u' S_{4}^{\mu \nu \rho \sigma}+t' S_{7}^{\mu \nu \rho \sigma}\right)+\frac{\left(\mz^2+\pt^2\right) }{4 \
\pt^2 }\left(\frac{S_{5}^{\mu \nu \rho \sigma}}{u'}+\frac{S_{6}^{\mu \nu \rho \sigma}}{t'}\right) \\
&+\frac{S_{8}^{\mu \nu \rho \sigma}}{\pt^2}-\frac{\left(\mz^2+\pt^2\right)}{2 \pt^2 s'}\left(S_{9}^{\mu \nu \rho \sigma}+S_{12}^{\mu \nu \rho \sigma}\right)+\frac{u' \
\left(-\mz^4+\pt^4-2 \pt^2 (s'-t'+u')\right) S_{10}^{\mu \nu \rho \sigma}}{2 \pt^2 \
\left(\mz^2+\pt^2\right) s' t'}  \nonumber \\
&-\frac{t' \
\left(\mz^4-\pt^4+2 \pt^2 (s'+t'-u')\right) S_{11}^{\mu \nu \rho \sigma}}{2 \pt^2 \
\left(\mz^2+\pt^2\right) s' \
u'}-\frac{\left(\mz^2+\pt^2\right)}{4 \pt^2 \
} \left(\frac{S_{13}^{\mu \nu \rho \sigma}}{u'}+\frac{S_{16}^{\mu \nu \rho \sigma}}{t'}\right) \nonumber \\
&-\frac{1}{2 \pt^2 s'}\left(u' S_{14}^{\mu \nu \rho \sigma}+t' \
S_{15}^{\mu \nu \rho \sigma}\right) +\frac{\left(\mz^2+\pt^2-s'+t'-u' \right) u' S_{18}^{\mu \nu \rho \sigma}}{\pt^2 \left(\mz^2+\pt^2\right) \
s' t'} \nonumber \\
&+\frac{t' \
\left(\mz^2+\pt^2-s'-t'+u'\right) \
S_{19}^{\mu \nu \rho \sigma}}{\pt^2 \left(\mz^2+\pt^2\right) s' \
u'} \nonumber \\
\mathcal{P}_{20}^{\mu \nu \rho \sigma} &=\frac{1}{2 \mz^2+\pt^2} \Biggl[ S_{3}^{\mu \nu \rho \sigma}-S_{2}^{\mu \nu \rho \sigma} -\frac{1}{\pt^2 s'} \left(u' S_{4}^{\mu \nu \rho \sigma}-t' S_{7}^{\mu \nu \rho \sigma} + u' S_{14}^{\mu \nu \rho \sigma}-t' S_{15}^{\mu \nu \rho \sigma}\right) \\
& -\frac{\left(\mz^2+\pt^2\right)}{2 \pt^2} \left(\frac{S_{5}^{\mu \nu \rho \sigma}}{u'}-\frac{S_{6}^{\mu \nu \rho \sigma}}{t'}+\frac{S_{13}^{\mu \nu \rho \sigma}}{u'}-\frac{S_{16}^{\mu \nu \rho \sigma}}{t'}\right) \nonumber \Biggr]
\end{align}

The form factors $\mathcal{A}_i$ associated to the above projectors can be expressed in terms of the form factors $f_i$ defined in eq.~(2.8) of ref.~\cite{Davies:2020lpf} via the following relations

\begin{align}
\mathcal{A}_{1}&=0  \label{eq:convers_eqs_init}\\
\mathcal{A}_{2}&= \frac{\pt^2}{\mz^2} \Biggl[ \frac{2 (t'-u') }{ \left(\mz^2+\pt^2\right)}\left(\pt^2 f_{8} -f_1\right)+ s'
   \pt^2\left(\frac{u'}{t'}f_{19}-\frac{t'}{u'}f_{18}\right) \\
   &+ 2 t' f_{4}-2 u' f_{5} + 2 t' f_{6} - 2 u' f_{7} +\frac{s' t'  }{ u'}f_{10}-\frac{s'  u'  }{ t'}f_{11}\Biggr] \nonumber \\
\mathcal{A}_{3}&= \frac{\pt^2}{\mz^2}\Biggl[f_{3}-f_{2}
+\frac{\left(\mz^2-\pt^2\right) s' }{2} \left(\frac{f_{7}}{t'}-\frac{f_{4}}{u'}\right) +u' f_{5} -t' f_{6} \Biggr] \\
\mathcal{A}_{4}&=0 \\
\mathcal{A}_{5}&= \frac{\pt^2}{\mz^2} \Biggl[\frac{2 (t'-u') }{ \left(\mz^2+\pt^2\right)} f_1 +\frac{s'  u'  }{ t'}f_{11} -\frac{s' t'  }{ u'}f_{10}\Biggr] \\
\mathcal{A}_{6}&=\frac{\pt^2}{\mz^2} \Biggl[f_{3}-f_{2}+t' f_{4} - u' f_{7} +\frac{\left(\mz^2-\pt^2\right) s'}{2 } \left(\frac{f_{6}}{u'}-\frac{f_5}{u'}\right)\Biggr] \\
\mathcal{A}_{7}&= \frac{\pt^2}{\mz^2} \Biggl[f_3-f_2 +\frac{\left(3
   s' \mz^2+\pt^2 s'-2 \left(u' \mz^2+t' \left(\mz^2+u'\right)\right)\right)  \pt^2}{4 \mz^2 } \left(\frac{f_{5}}{t'}-\frac{f_{6}}{u'}\right) \Biggr] \\
\mathcal{A}_{8}&= \frac{\pt^2}{\mz^2} \Biggl[f_3-f_2 \frac{\left(3
   s' \mz^2+\pt^2 s'-2 \left(u' \mz^2+t' \left(\mz^2+u'\right)\right)\right)  \pt^2}{4 \mz^2 } \left(\frac{f_{4}}{u'}-\frac{f_{7}}{t'}\right) \Biggr] \\
	\mathcal{A}_{9}&=\frac{1}{\mz^4}\Biggl[-2 \pt^4 \left(t' f_4 + u' f_5+t' f_6+u' f_7\right) - \left(\mz^2+\pt^2\right) \pt^2 \left(f_2+f_3\right) \\
	& -\frac{\left( s' (\mz^2 + \pt^2) (\mz^2 -s') + 2 \mz^2 (t'^2 + u'^2) \right)}{s'} \left(\pt^2 f_8 - f_1 \right)  -s' \left(\mz^4+\pt^4\right) \left(f_{9}-\pt^2 f_{20}\right) \nonumber \\
   & +\frac{2 \left(u' \mz^2+t' \left(\mz^2+u'\right)\right)}{s'} \left(\pt^2\left(t'^2 f_{18} +u'^2 f_{19}\right)-\left(t'^2 f_{10} + u'^2 f_{11}\right)\right)\Biggr] \nonumber \\
	\mathcal{A}_{10}&= \frac{2 \pt^2}{(\mz^4+\pt^4)} \Biggl[s' \pt^2\left( \pt^2\left(\frac{t'}{u'} f_{18} +\frac{u'}{t'} f_{19}\right)-\left(\frac{t'}{u'} f_{10} +\frac{u'}{t'} f_{11}\right)\right)- 2 \pt^2 \left(t' f_{4}+u' f_{5}+t' f_{6} + u' f_{7}\right)  \\
   &+\frac{ \left(\left(s'^2-t'^2-u'^2\right) \mz^2+\pt^2 \left(s'^2+t'^2+u'^2\right)\right)}{\left(\mz^2+\pt^2\right) s'}\left(f_{8}
   \pt^2- f_{1}\right) - \left(\mz^2+\pt^2\right) \left(f_2+f_3\right)\Biggr] \nonumber  \\
   \mathcal{A}_{11}&= \frac{1}{\mz^4}  \Biggl[s' \left(\mz^4 + \pt^4\right) f_{9} + \frac{2 \left(u' \mz^2+t'
   \left(\mz^2+u'\right)\right)}{s'}  \left(t'^2 f_{10} + u'^2 f_{11}\right) \\
  &-\frac{\left(2 u'^2 \mz^2+s' \left(\mz^2+u'\right) \left( s' \mz^2 +2 t'^2 + \pt^2 s' \right) \right)}{ s'} f_{1}\Biggr] \nonumber 
\end{align}
\begin{align}
\mathcal{A}_{12}&=\frac{2 \pt^2}{\left(\mz^4+\pt^4\right)} \Biggl[\frac{\left(\left(s'^2-t'^2-u'^2\right) \mz^2+\pt^2
   \left(s'^2+t'^2+u'^2\right)\right)  }{\left(\mz^2+\pt^2\right)  s'}f_{1} + \pt^2 s' \left(\frac{t'}{u'}f_{10}+ \frac{u'}{t'} f_{11}\right)\Biggr] \\
   \mathcal{A}_{13}&=\frac{\pt^2}{\mz^2} \Biggl[\frac{\left(\mz^2-\pt^2\right) s'}{2} \left(\frac{f_5}{t'} + \frac{f_6}{u'}\right)-\left(f_2 + f_3 + t' f_4 + u' f_7\right)\Biggr] \\
   \mathcal{A}_{14}&= -\frac{\pt^2}{\mz^2} \Biggl[f_2 +f_3 + \pt^2 s' \left(\frac{f_5}{t'} + \frac{f_6}{u'}\right)\Biggr] \\
   \mathcal{A}_{15} &= \frac{\pt^2}{2 \mz^4} \Biggl[\left(\mz^2+\pt^2\right) \left(f_2+f_3 + t' f_4 + u' f_7\right) - \left(\mz^2 -\pt^2\right) \left(u' f_5 + t' f_6\right)\Biggr] \\
   \mathcal{A}_{16}&= \frac{2 \pt^2}{\mz^2 + \pt^2} \Biggl[f_2 + f_3 + \frac{2 \pt^2}{\mz^2 + \pt^2} \left(u' f_5 +t' f_6\right)\Biggr] \\
\mathcal{A}_{17} &= \frac{2 \mz^2}{2 \mz^2 +  \pt^2} \Biggl[\frac{\left( s'^2\left(t'^2 + u'^2\right)-\left(t'^2-u'^2\right)^2 \right)}{s'^3 \left(\mz^2 + \pt^2\right)} f_1 +f_2 + f_3\Biggr] \\
\mathcal{A}_{18}&= \frac{\left( \mz^2 \left(2 s'^2 -t'^2 -u'^2\right) + s' (s'(\pt^2-s')+t'^2 +u'^2) \right)}{\mz^2 s' \left( \mz^2 +\pt^2 \right)} \left(\pt^2 f_8 -f_1\right) 
\label{eq:convers_eqs_end}
\end{align}
\newpage

\end{appendletterA}

\begin{appendletterB}
\section{Analytical Results}
\label{app:analytzz}


\paragraph{Results for Double-Triangle Diagrams}
 With reference to eq.~(\ref{eq:nlosplitzz}), we present here  the exact results for the  
double-triangle contributions to the $\mathcal{A}^{(1, \bowtie)}_i$ with $i=1,\dots, 18$. We keep the dependence of the final result on the mass of the bottom quark, $m_b$.  We find
\begin{align}
\mathcal{A}^{(1, \bowtie)}_1 &=0 \\
\mathcal{A}^{(1, \bowtie)}_2 &= \frac{\pt \left(\left(\mz^2+2 t'\right) \
\Delta(t')^2- (t' \leftrightarrow u') \right)}{32 \sqrt{2} \mz \
\left(\mz^2+\pt^2\right)} \\
\mathcal{A}^{(1, \bowtie)}_3 &= -\frac{\pt^3 \left(u' \Delta(t')^2 -(t' \leftrightarrow u') \right) }{16 \
\sqrt{2} \mz \left(\mz^2+\pt^2\right)^2} \\
\mathcal{A}^{(1, \bowtie)}_4 &= 0 \\
\mathcal{A}^{(1, \bowtie)}_5 &= -\frac{\pt \
\left(t' \
\Delta(t')^2 -(t' \leftrightarrow u')\right)}{16 \sqrt{2} \mz \
\left(\mz^2+\pt^2\right)} \\
\mathcal{A}^{(1, \bowtie)}_6 &= -\frac{\pt \left(\left(\mz^2+2 t'\right) u'^2 \
\Delta(t')^2 -(t' \leftrightarrow u') \right)}{16 \sqrt{2} \mz \
\left(\mz^2+\pt^2\right)^2 s'} \\
\mathcal{A}^{(1, \bowtie)}_7 &= \frac{\left(s' \
\pt^2+t'^2\right) u'^2 \
\Delta(t')^2-(t' \leftrightarrow u')}{16 \sqrt{2} \
\left(\mz^2+\pt^2\right)^2 s'^2} \\
\mathcal{A}^{(1, \bowtie)}_8 &= \frac{u' \left(t' (\mz^2+\pt^2)-2 s' \pt^2\right) \Delta(t')^2-(t' \leftrightarrow u')}{32 \sqrt{2} \
\left(\mz^2+\pt^2\right)^2 \
s'} \\
\mathcal{A}^{(1, \bowtie)}_9 &= \frac{\left(\mz^2 - \pt^2\right) \left(\left(\mz^2+2 t'\right) \
\Delta(t')^2+ (t' \leftrightarrow u')\right)}{64 \mz^2 \
\sqrt{\mz^4+\pt^4} } \\
\mathcal{A}^{(1, \bowtie)}_{10} &= -\frac{\pt^2 \
\left(\left(\mz^2+2 t'\right) \
\Delta(t')^2+(t' \leftrightarrow u')\right)}{32 \left(\mz^2+\pt^2\right) \
\sqrt{\mz^4+\pt^4}} \\
\mathcal{A}^{(1, \bowtie)}_{11} &= \frac{u' \left(\mz^2+2 \
t'\right) \left(2 s' \
\mz^2+\left(\mz^2+\pt^2\right) (t'+2 \
u')\right) \Delta(t')^2+(t' \leftrightarrow u')}{32 \mz^2 \
\left(\mz^2+\pt^2\right) \sqrt{\mz^4+\pt^4} \
s'} \\
\mathcal{A}^{(1, \bowtie)}_{12} &= -\frac{\left(s' \mz^4+4 \pt^2 s' \
\mz^2+3 \pt^4 s'+4 \pt^2 \
\left(t'^2+u'^2\right)\right) \
\left(\Delta(t')^2+\Delta(u')^2\right)}{64 \
\left(\mz^2+\pt^2\right) \sqrt{\mz^4+\pt^4} \
s'} \\
\mathcal{A}^{(1, \bowtie)}_{13} &= \frac{\pt \left(\left(\mz^2+2 t'\right) \
u'^2 \Delta(t')^2+(t' \leftrightarrow u')\right)}{16 \sqrt{2} \mz \
\left(\mz^2+\pt^2\right)^2 s'} 
\end{align}

\begin{align}
\mathcal{A}^{(1, \bowtie)}_{14} &= -\frac{\left(s' \
\pt^2+t'^2\right) u'^2 \
\Delta(t')^2+(t' \leftrightarrow u')}{64 \sqrt{2} \
\left(\mz^2+\pt^2\right)^2 s'^2} \\
\mathcal{A}^{(1, \bowtie)}_{15} &= -\frac{\pt \
\left(\left(\mz^2+2 t'\right) u'^2 \
\Delta(t')^2+(t' \leftrightarrow u')\right)}{16 \sqrt{2} \mz \
\left(\mz^2+\pt^2\right)^2 s'} \\
\mathcal{A}^{(1, \bowtie)}_{16} &= \frac{\left(s' \
\pt^2+t'^2\right) u'^2 \
\Delta(t')^2+(t' \leftrightarrow u')}{16 \sqrt{2} \
\left(\mz^2+\pt^2\right)^2 s'^2} \\
\mathcal{A}^{(1, \bowtie)}_{17} &=0 \\
\mathcal{A}^{(1, \bowtie)}_{18} &= \frac{1}{64} \left(\left(\frac{s' \
\pt^2}{t'^2}+1\right) \
\Delta(t')^2+(t' \leftrightarrow u')\right)
\end{align}
where
\begin{equation}
\Delta(x) = F(x,m_t) - F(x,m_b)
\end{equation}
and 
\begin{align}
F(x,M) &= \frac{\mz^2}{x} \left[ B_0(2x + \mz^2,M^2,M^2) - B_0(\mz^2,M^2,M^2) \right] \\
&+ 4 M^2 C_0 (0,\mz^2,2x+\mz^2,M^2,M^2,M^2) + 2 \nonumber
\end{align}

\newpage
\paragraph{Results for Triangle Diagrams at NLO} 
 With reference to eq.~(\ref{eq:nlosplitzz}), we present here  the exact results for the  
two-loop one-particle-irreducible triangle contributions to the NLO form factors. We obtain

\begin{align}
\mathcal{A}^{(1, \triangle)}_1 &=0 \\
\mathcal{A}^{(1, \triangle)}_2 &= \frac{\pt^2  (\partt-\partu)}{\mz^2(\pt^2+\mz^2)} \frac{\parts}{\parts - m_H^2} \left(C_F \mathcal{F}_{1/2}^{(2l)}+C_A \mathcal{G}_{1/2}^{(2l,CA)} \right) \\
\mathcal{A}^{(1, \triangle)}_3 &=0 \\
\mathcal{A}^{(1, \triangle)}_4 &=0 \\
\mathcal{A}^{(1, \triangle)}_5 &= - \frac{\pt^2  (\partt-\partu)}{\mz^2(\pt^2+\mz^2)} \frac{\parts}{\parts - m_H^2} \left(C_F \mathcal{F}_{1/2}^{(2l)}+C_A \mathcal{G}_{1/2}^{(2l,CA)} \right) \\
\mathcal{A}^{(1, \triangle)}_6 &=0 \\
\mathcal{A}^{(1, \triangle)}_7 &=0 \\
\mathcal{A}^{(1, \triangle)}_8 &=0 \\
\mathcal{A}^{(1, \triangle)}_9 &=\frac{( \parts(\pt^2-\mz^2) +2\mz^2( \pt^2  + \mz^2)}{2 \mz^4}  \frac{\parts}{\parts - m_H^2} \left(C_F \mathcal{F}_{1/2}^{(2l)}+C_A \mathcal{G}_{1/2}^{(2l,CA)} \right)\\
\mathcal{A}^{(1, \triangle)}_{10} &=\frac{2 \pt^2(\pt^2 \parts+\mz^4-\pt^4)}{(\pt^2+ \mz^2)(\pt^4+ \mz^4)}  \frac{\parts}{\parts - m_H^2} \left(C_F \mathcal{F}_{1/2}^{(2l)}+C_A \mathcal{G}_{1/2}^{(2l,CA)} \right)\\
\mathcal{A}^{(1, \triangle)}_{11} &=-\frac{( \parts(\pt^2-\mz^2) +2\mz^2( \pt^2  + \mz^2)}{2 \mz^4}  \frac{\parts}{\parts - m_H^2} \left(C_F \mathcal{F}_{1/2}^{(2l)}+C_A \mathcal{G}_{1/2}^{(2l,CA)} \right)\\
\mathcal{A}^{(1, \triangle)}_{12} &=-\frac{2 \pt^2(\pt^2 \parts+\mz^4-\pt^4)}{(\pt^2+ \mz^2)(\pt^4+ \mz^4)}  \frac{\parts}{\parts - m_H^2} \left(C_F \mathcal{F}_{1/2}^{(2l)}+C_A \mathcal{G}_{1/2}^{(2l,CA)} \right)\\
\mathcal{A}^{(1, \triangle)}_{13} &=0 \\
\mathcal{A}^{(1, \triangle)}_{14} &=0 \\
\mathcal{A}^{(1, \triangle)}_{15} &=0 \\
\mathcal{A}^{(1, \triangle)}_{16} &=0 \\
\mathcal{A}^{(1, \triangle)}_{17} &=-\frac{2 \mz^2}{\pt^2 + 2 \mz^2}\frac{\parts}{\parts - m_H^2} \left(C_F \mathcal{F}_{1/2}^{(2l)}+C_A \mathcal{G}_{1/2}^{(2l,CA)} \right) \\
\mathcal{A}^{(1, \triangle)}_{18} &=\frac{\parts}{\parts - m_H^2} \left(C_F \mathcal{F}_{1/2}^{(2l)}+C_A \mathcal{G}_{1/2}^{(2l,CA)} \right)
\end{align}
where the functions $\mathcal{F}_{1/2}^{(2l)}$ and $\mathcal{G}_{1/2}^{(2l,CA)}$ are defined in eqs.~(2.11) and (3.8) in ref.~\cite{Aglietti:2006tp}.

\end{appendletterB}

\bibliographystyle{utphys}
\bibliography{DGV}

\end{document}